\documentclass{aa}  

\usepackage{graphicx}
\usepackage{multirow}
\usepackage{amssymb}
\usepackage{xcolor}
\usepackage{txfonts}
\usepackage{hyperref}
\begin{document} 

   \title{Type IIP supernova SN\,2016X in radio frequencies}
   
   \subtitle{}

   \author{R. Ruiz-Carmona
          \inst{1}\fnmsep\thanks{\email{ Roque.RuizCarmon@mail.huji.ac.il}},
          I. Sfaradi\inst{1}
          \and
          A. Horesh\inst{1}
          }

   \institute{Racah Institute of Physics. The Hebrew University of Jerusalem. Jerusalem 91904, Israel}

   \date{Received September 15, 1996; accepted March 16, 1997}

 
  \abstract
   {The study of radio emission from core-collapse supernovae (SNe) probes the interaction of the ejecta with the circumstellar medium (CSM) and reveals details of the mass-loss history of the progenitor.}
   {We report observations of the type IIP supernova SN\,2016X during the plateau phase, at ages between 21 and 75 days, obtained with the Karl G. Jansky Very Large Array (VLA) radio observatory.}
   {We modelled the radio spectra as self-absorbed synchrotron emission, and we characterised the shockwave and the mass-loss rate of the progenitor. We also combined our results with previously reported X-ray observations to verify the energy equipartition assumption.}
   {The properties of the shockwave are comparable to other type IIP supernovae. The shockwave expands according to a self-similar law $R \propto t^m$ with $m=0.76 \pm 0.08$, which is notably different from a constant expansion. The corresponding shock velocities are approximately 10700 -- 8000\,km\,s$^{-1}$ during the time of our observations. The constant mass-loss rate of the progenitor is $\dot{M}=$\,(7.8 $\pm$\, 0.9)\,$\times\, 10^{-7}\, \alpha^{-8/19}\, (\epsilon_B/0.1)^{-1} M_{\odot}$\, yr$^{-1}$, for an assumed wind velocity of 10\,km\,s$^{-1}$. We observe spectral steepening in the optically thin regime at the earlier epochs, and we demonstrate that it is caused by electron cooling via the inverse Compton effect. We show that the shockwave is characterised by a moderate deviation from energy equipartition by a factor of $\epsilon_e / \epsilon_B \approx 28$, being the second type IIP supernova to show such a feature.}
   {}


   \maketitle
%

\section{Introduction}

Massive stars experience mass loss via stellar winds or due to eruptive events that can produce enhanced mass-loss rates towards the end of their lives \citep{10.1086/513145, 10.1038/nature05825, 10.1088/0004-6256/139/4/1451, 10.1088/0004-637x/797/2/107, 10.1093/mnras/stu163, 10.1088/0004-637x/789/2/104, Bruch20, 10.3847/1538-4357/abd032}. Mass loss enriches the circumstellar medium (CSM), affects the subsequent evolution of the star, and influences the phenomenology of some supernovae depending on the amount of mass lost at the moment of explosion (e.g. \citealt{10.1088/0004-6256/137/3/3558, 2010ASPC..425..279V}).

These massive stars die in core-collapse supernova (SN) explosions, once their cores cannot be supported by thermonuclear reactions (see \citealt{10.1017/pasa.2015.17} for a review). SN explosions eject large amounts of mass at high velocities, that collide with the pre-existing CSM and produce shocks, which in turn accelerate particles to relativistic velocities and enhance the local magnetic field. As a result, synchrotron radiation is generated, which can be prominent in radio and X-ray wavelengths \citep{10.1086/160167}. Conversely, this emission can be used to constrain the mass-loss history of the progenitor, and this provides key information on the latest stages of stellar evolution. 

The radio emission from SNe has been theoretically modelled \citep{10.1086/160167, 10.1086/305676, 10.1146/annurev.astro.40.060401.093744, 10.1086/500528}, and observationally studied in several cases (e.g. \citealt{1990ApJ...364..611W, 10.1086/305676, 2003ApJ...592..900S, 2004MNRAS.349.1093R, 2005ApJ...621..908S, 10.1088/0004-637x/778/1/63, 2014ApJ...780...21M, 2014ApJ...782...42C,  10.3847/1538-4357/abccd9}). Radio observations at early phases after the SN explosion, in combination with X-ray observations, have revealed properties of the CSM shock, provided estimates of the mass-loss rate of the progenitor, and constrained the fraction of post-shock energy converted to particle acceleration and magnetic field amplification (e.g. \citealt{10.1088/0004-637x/725/1/922, 10.1088/0004-637x/778/1/63, 10.1093/mnras/stt1645, 2014ApJ...780...21M, 10.1088/0004-637x/782/1/30, 10.3847/0004-637x/818/2/111, Horesh_2020}).

Type IIP SNe (SNe IIP) are the most abundant SN explosions; over half of all core-collapse SNe in several volume-limited samples are SNe IIP (\citealt{10.1146/annurev-astro-082708-101737} and references therein). SNe IIP exhibit a long plateau phase in the optical light curve, associated with the hydrogen envelope of the progenitor star. The progenitors of SNe IIP are red supergiant stars (RSGs; \citealt{10.1086/500528}). RSGs are characterised by masses in the range  8--25 M$_\odot$ and radii in the range  500--1500 R$_\odot$ (\citealt{10.1086/375341, 10.1051/0004-6361:20041095}, \citealt{1976ApJ...207..872C, 1980ApJ...237..541A}), although pre-explosion imaging has only identified progenitors towards the lower mass so far \citep{2003PASP..115....1V, 2005PASP..117..121L, 2005MNRAS.360..288M}. RSGs are believed to sustain slow winds that drive moderate mass loss at rates  $\dot{M} = 10^{-6}$--$10^{-4}\, \rm{M}_{\odot}$ with a wind velocity of approximately $w \approx 10$\,km\,s$^{-1}$ \citep{10.1146/annurev-astro-081913-040025}. However, the mass loss from these stars has been measured empirically in a small number of cases.

In this paper we present radio observations of the type IIP supernova SN\,2016X. We model the radio spectra with the standard SN--CSM interaction model \citep{10.1086/159460} and derive the properties of the shock and the mass-loss rate of the progenitor. 

The paper is structured as follows. In Section~\ref{sec:sn2016x} we summarise the previously published results on SN\,2016X. In Section~\ref{sec:observations} we present our radio observations in detail.   In Section~\ref{sec:analysis} we then describe the SN--CSM interaction model and analyse the radio spectra to characterise the shock and calculate the mass-loss rate of the progenitor. Electron cooling and its effects are discussed in Section~\ref{sec:cooling}, while in Section~\ref{sec:x-ray} we discuss the possible deviation from the energy equipartition assumption. Finally, we discuss our results and summarise our conclusions in Section~\ref{sec:discussion}.

\section{SN\,2016X}
\label{sec:sn2016x}

SN\,2016X was discovered on 20.6 January 2016 by the All Sky Automated Survey for SuperNovae (ASASSN-16at; \citealt{2016ATel.8566....1B}), and classified as a SN IIP using optical spectroscopy acquired within the first three days after discovery \citep{2016ATel.8567....1H, 2016ATel.8584....1Z}. The adopted time of the explosion is 18.9 January 2016 (MJD = 57405.92 $\pm$ 0.57; \citealt{10.1093/mnras/sty066}), based on a non-detection ($V>18.0$\,mag) on 18.4 January and a first detection ($V\sim16.6$\,mag) on 19.5 January. The host galaxy, UGC 08041, is at a distance of 15.2\,Mpc (Tully-Fisher distance, \citealt{10.1093/mnras/stu1450}), and we adopt this as the distance to SN\,2016X.

\citet{10.1093/mnras/sty066} and \citet{10.3847/2041-8213/ab0558} present high-cadence optical and ultraviolet light curves and spectra from the early phases after explosion until the nebular phase. The radius of the progenitor star, as revealed by the rise time, is in the range  860 -- 990\,R$_\odot$ \citep{10.1051/0004-6361/201525868}. Using mass-radius relations for RSGs, the mass of the progenitor is estimated to be 18.5--19.7\,M$_\odot$. The quasi-bolometric luminosity is comparable to other SNe IIP, but it declines quickly during the plateau phase. The study of absorption features in the spectral lines H$\alpha$, H$\beta$, and Fe\,{\sc ii} 5169\AA, 5018\AA, and 4924\AA\, shows that photospheric velocities are high compared to other SNe IIP such as SN\,1999em and SN\,2005cs, but  are comparable to SN\,2013ej and SN\,2014cx \citep{10.1086/324785, 10.1111/j.1365-2966.2009.14505.x, 10.1088/0004-637x/807/1/59, 10.3847/0004-637x/832/2/139}, evolving rapidly from $\sim$9000\,km\,s$^{-1}$ at $t$ = 20\,days to $\sim$6000\,km\,s$^{-1}$ at $t$ = 80\,days for H$\alpha$.

SN\,2016X was also observed in X-rays with the Neil Gehrels Swift Observatory and the Chandra X-ray Observatory in the 0.3--10\,keV energy band. The observed X-ray luminosity was $L_X\sim$\,2.5 $\times$\,$10^{39}$\,erg\,s$^{-1}$ at $t\sim$2--5\,days, and declined by a factor of 5 by $t\sim$\,19\,days \citep{10.3847/2041-8213/ab0558}. The count rate is too low to perform any spectral analysis on the X-ray observations.

Finally, it is noteworthy that SN\,2016X showed double-peaked Balmer lines with strong asymmetries during the nebular phase. This phenomenology has been associated with bipolar ejecta or aspherical distribution of the CSM \citep{10.3847/2041-8213/ab0558, 10.1093/mnras/stz2716}, and possibly with the formation of dust in the ejecta \citep{10.1111/j.1365-2966.2011.19128.x, 10.3847/2041-8213/ab0558}.


\section{Observations}
\label{sec:observations}

We observed SN\,2016X ($\alpha$ = 12$^{\rm{h}}$\,55$^{\rm{m}}$\,15$^{\rm{s}}.61$, $\delta$ = 12$^{\circ}$\,05'\,59''.35, J2000) with the National Radio Astronomy Observatory (NRAO) Karl G. Jansky Very Large Array (VLA) under the programme 15B-395 (PI A. Horesh). Observations were acquired at four epochs: 9.50 February, 15.47 February,  2.56 March, and 3.84 April 2016, which correspond to 21.58, 27.55, 43.54, and 75.42 days after the explosion, adopting the ephemeris of the explosion by \citet{10.3847/2041-8213/ab0558}. Observations were carried out in bands S (2--4\,GHz), C (4--8\,GHz), X (8--12\,GHz), Ku (12--18\,GHz), and K (18--26\,GHz) in the three later  epochs, while only C- and K-band observations were obtained in the first epoch. The antenna array was in configuration C. 

\begin{figure*}[ht!]
  \begin{center}
    \includegraphics[width=0.9\linewidth]{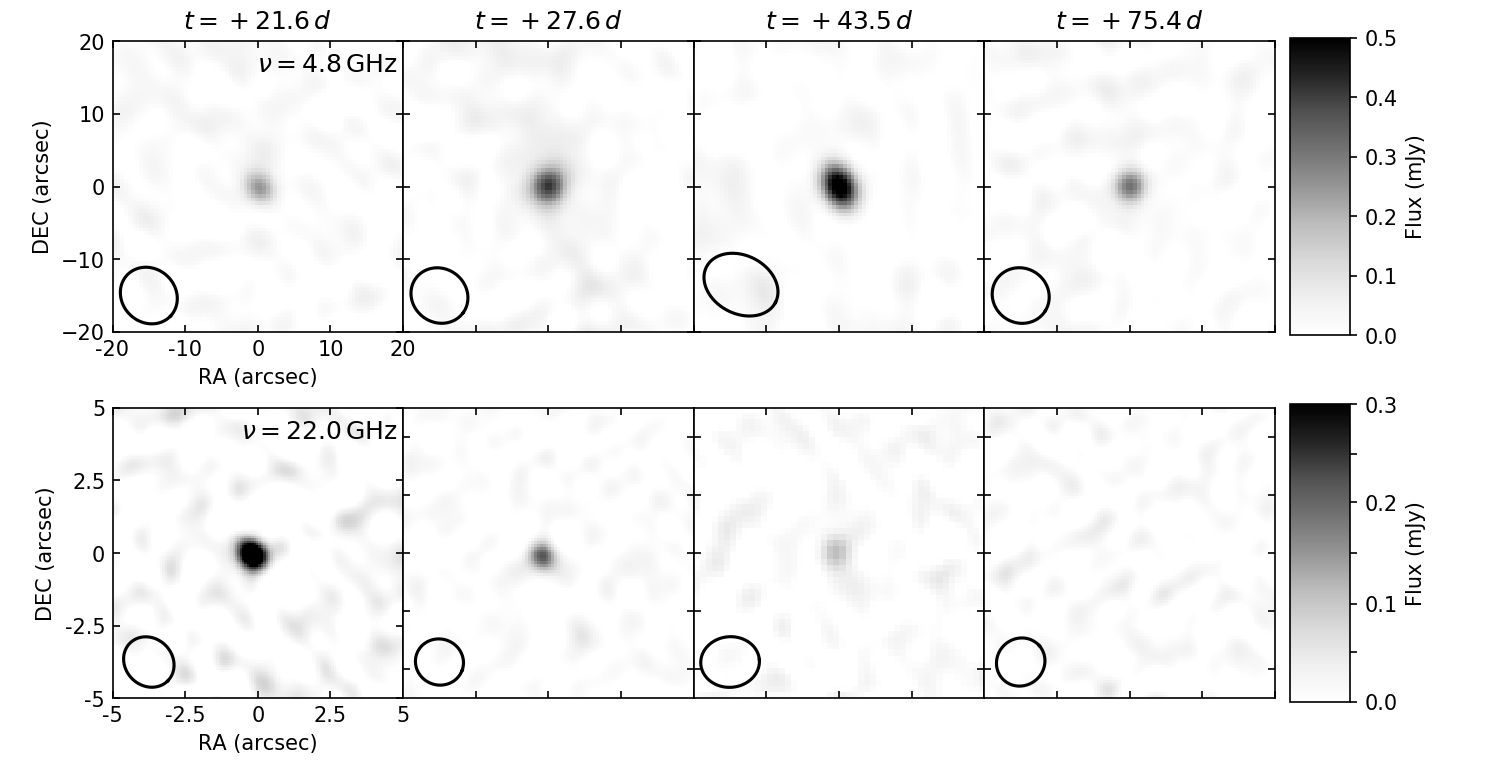}
    \caption{Images of SN\,2016X at the four epochs of our observations, obtained at 4.8\,GHz (top row) and 22.0\,GHz (bottom row). The flux on the images is indicated by the colour bars on the right, and the beam size is represented at the bottom left of every image.} 
    \label{fig:images}
  \end{center}
\end{figure*}

We reduced the observational data with the Common Astronomy Software Applications package ({\sc casa}; \citealt{2007ASPC..376..127M}). The reduction was done manually, as additional flagging was needed in most bands. The radio sources J\,1246--0730 and 3C\,286 were used as phase and flux calibrators, respectively. We imaged the field of SN\,2016X in several sub-bands depending on the signal-to-noise ratio of the detection and the predominance of radio frequency interference (RFI) using the task {\sc clean} interactively. We extracted the flux of the source by fitting a two-dimensional Gaussian with the task {\sc imfit}, and we added in quadrature the contribution of three factors to estimate the error bars on the flux: (a) the error provided by the fitting routine, (b) the RMS of the image estimated in an empty region around the source, and (c) 10\% of the peak flux to account for calibration errors. We adopted three times the RMS of the image as a 3$\sigma$ upper limit on the flux for non-detections. In the S-band images, SN\,2016X appears embedded in the side lobe of a bright source in the field. We do not use the S-band data in this work since the extracted flux in that band is artificially high. A summary of the observations is presented in Table~\ref{tab:obs}, and images at 4.8 and 22.0\,GHz are represented in Figure~\ref{fig:images}.

\begin{table}
\caption[]{Summary of radio observations of SN\,2016X.}
\label{tab:obs}
\begin{tabular}{c c c c}
    \hline \noalign{\smallskip}
    Epoch & Central Frequency & Bandwidth & Flux \\
    (days) & (GHz) & (GHz) & ($\mu$Jy)\\

    \noalign{\smallskip} \hline \noalign{\smallskip}
    +21.6   &   4.8 & 1.0 & 220 $\pm$ 41 \\
    +21.6   &   7.4 & 1.0 & 659 $\pm$ 72 \\
    +21.6   &  18.8 & 1.7 & 659 $\pm$ 77 \\
    +21.6   &  22.0 & 1.5 & 454 $\pm$ 67 \\
    +21.6   &  23.6 & 1.6 & 519 $\pm$ 79 \\
    +21.6   &  25.2 & 1.7 & 495 $\pm$ 61 \\
    \noalign{\smallskip} 
    +27.6   &   4.8 & 1.0 & 404 $\pm$ 47 \\
    +27.6   &   7.4 & 1.0 & 739 $\pm$ 73 \\
    +27.6   &   8.4 & 0.8 & 735 $\pm$ 81 \\
    +27.6   &   9.1 & 0.8 & 631 $\pm$ 78 \\
    +27.6   &   9.9 & 0.9 & 628 $\pm$ 66 \\
    +27.6   &  10.9 & 1.2 & 575 $\pm$ 68 \\
    +27.6   &  13.3 & 1.0 & 457 $\pm$ 53 \\
    +27.6   &  14.3 & 1.0 & 455 $\pm$ 56 \\
    +27.6   &  15.3 & 1.0 & 402 $\pm$ 52 \\
    +27.6   &  16.3 & 1.0 & 394 $\pm$ 54 \\
    +27.6   &  17.4 & 1.3 & 341 $\pm$ 48 \\
    +27.6   &  22.0 & 8.0 & 233 $\pm$ 46 \\
    \noalign{\smallskip}
    +43.5   &   4.8 & 1.0 & 537 $\pm$ 65 \\
    +43.5   &   7.4 & 1.0 & 475 $\pm$ 57 \\
    +43.5   &   8.4 & 0.8 & 443 $\pm$ 58 \\
    +43.5   &   9.1 & 0.8 & 437 $\pm$ 58 \\
    +43.5   &   9.9 & 0.9 & 345 $\pm$ 42 \\
    +43.5   &  10.9 & 1.2 & 298 $\pm$ 44 \\
    +43.5   &  13.6 & 1.7 & 220 $\pm$ 32 \\
    +43.5   &  15.4 & 1.7 & 211 $\pm$ 33 \\
    +43.5   &  17.1 & 1.8 & 174 $\pm$ 32 \\
    +43.5   &  22.0 & 8.0 & 145 $\pm$ 26 \\
    \noalign{\smallskip}
    +75.4   &   4.8 & 1.0 & 284 $\pm$ 45 \\
    +75.4   &   7.4 & 1.0 & 193 $\pm$ 33 \\
    +75.4   &   8.4 & 0.8 & 194 $\pm$ 41 \\
    +75.4   &   9.1 & 0.8 & 163 $\pm$ 27 \\
    +75.4   &   9.9 & 0.9 & 165 $\pm$ 31 \\
    +75.4   &  10.9 & 1.2 & 162 $\pm$ 29 \\
    +75.4   &  15.4 & 5.3 &  93 $\pm$ 16 \\
    +75.4   &  22.0 & 8.0 & $<90$ $\pm$ 30 \\
    \noalign{\smallskip} \hline \noalign{\smallskip}
    \end{tabular}

{\small The epoch of the observations is estimated from the explosion time (MJD = 57405.92 $\pm$ 0.57; \citealt{10.1093/mnras/sty066}). The flux is calculated fitting a 2D Gaussian profile to the images of the sub-bands specified by their central frequency and bandwidth. The error bars are calculated adding in quadrature the error of the fitting routine, the RMS of the image, and 10\% of the peak flux in order to account for calibration errors. }
\end{table}

\section{Analysis of radio observations}
\label{sec:analysis}

We use the SN--CSM interaction model by \citet{10.1086/159460} to study the radio emission from SN\,2016X. We make the following assumptions in our analysis: (i) we model the radio flux as self-absorbed synchrotron radiation, as described in \citet{10.1086/305676} and \citet{10.1086/500528}, although we note that other absorption mechanisms such as free-free absorption by ionised gas may also play a role in the case of SNe IIP (see Section~\ref{ssec:ffa}); (ii) we assume that the density profile of the CSM decreases with radius as $\rho_{\rm{CSM}} \propto R^{-2}$; (iii) we assume that only half of the spherical volume inside the radius $R$ is contributing to the radio emission at all times (i.e. the filling factor is $f=0.5$); (iv) we do not include the effects of electron cooling in our modelling; and (v) we assume that there is energy equipartition between particles and fields, that is, the fraction of the post-shock energy converted into particle acceleration, $\epsilon_e$, and the fraction of the post-shock energy converted into magnetic field amplification, $\epsilon_B$, are equal and approximately 0.1. Therefore, the equipartition parameter $\alpha$ becomes unity ($\alpha \equiv \epsilon_e / \epsilon_B$ = 1). We note that deviations from the assumptions above will affect the shockwave and mass-loss estimates, as we investigate in Sections~\ref{sec:cooling} and~\ref{sec:x-ray}.

The energy distribution of the relativistic electrons follows a power law, $N(E)=N_0 E^{-\gamma}$, where $E$ is the electron energy and $N_0$ is a constant. Under the energy equipartition assumption we can relate the constant $N_0$ and the magnetic field strength, $B$,
\begin{equation}
    N_0 = \frac{1}{8\pi} \alpha B^2 (\gamma-2) E_l^{\gamma-2},
\end{equation}where $E_l = 0.51$\, MeV is the rest mass energy of the electron.

The complete expression that describes the synchrotron emission from SNe is given in Equation 1 of \citet{10.1086/305676}. The observable flux in the optically thick regime is

\begin{equation}
    \label{eq:thick}
    F_{\nu} =  \frac{\pi c_5}{c_6}\, \frac{R^2}{D^2}\, B^{-1/2}\, \left( \frac{\nu}{2 c_1} \right)^{5/2},
\end{equation}and in the optically thin regime yields

\begin{equation}
    \label{eq:thin}
    F_{\nu} = \frac{1}{6}\, \alpha f c_5\, (\gamma-2)\, E_l^{\gamma-2} \, \frac{R^2}{D^2} \, B^{(\gamma+5)/2} \, \left( \frac{\nu}{2 c_1} \right)^{-(\gamma-1)/2}.
\end{equation}Here $c_1$ is a constant \citep{10.1086/305676}; $c_5$ and $c_6$ are functions of the electron energy distribution power-law index $\gamma$ \citep{1970ranp.book.....P}; $D$ is the distance to the supernova; $R$ is the radius of the forward shock; $B$ is the magnetic field strength; and $f$ is the filling factor. The spectral slope in the optically thin part of the spectrum, $\beta$, is a function of the electron energy distribution power-law index: $\beta = -(\gamma-1)/2$.

It can be seen from Equations~\ref{eq:thick} and \ref{eq:thin} that the transition from optically thick to optically thin emission creates a peak in the spectrum. By characterising the observed spectral peak by its frequency, $\nu_{\rm{peak}}$, and its flux, $F_{\rm{peak}}$, we can obtain the following expressions for the radius of the shock and the magnetic field strength (in cgs units, see \citealt{10.1086/305676}):

\begin{equation}
    \label{eq:r}
    R = \frac {2 c_1}{\nu_{\rm{peak}}} \, \left( \frac{6\pi\, c_6^{\gamma+5}\, ( 0.5\, {F}_{\rm{peak}}\, D^2) ^{\gamma+6}\,  }{c_5^{\gamma+6} \alpha\, f\, (\gamma-2)\, E_l^{\gamma-2} } \right)^{1/(2\gamma + 13)}
,\end{equation}

\begin{equation}
    \label{eq:b}
    B = \frac{\nu_{\rm{peak}}}{2 c_1} \, \left( \frac{706\, c_5}{ \left(\alpha\, f\, (\gamma-2)\, E_l^{\gamma-2} \right)^2\, c_6^3\, F_{\rm{peak}}\, D^2} \right)^{2/(2\gamma + 13)} 
.\end{equation}

The mass-loss rate of the progenitor, $\dot{M}$, divided by the wind velocity, $w$, can be calculated using the fact that the magnetic energy density is a fraction $\epsilon_B$ of the post-shock energy,

\begin{equation}
    \label{eq:b_scaling}
    \frac{B^2}{8\pi} = \frac{9}{8}\, \epsilon_B\, \rho_{\rm{CSM}}\, V_{\rm{sh}}^2
,\end{equation}
where $\rho_{\rm{CSM}} = \dot{M} / 4\pi\,w\,R^2$ is the density of the CSM, and $V_{\rm{sh}}$ is the expansion velocity of the shock. Assuming constant expansion velocity ($ V_{\rm{sh}} = R / t$), and using Equation~\ref{eq:b} and ~\ref{eq:b_scaling}, we obtain

\begin{equation}
    \label{eq:m2w}
    \frac{\dot{M}}{w} = \frac{1}{9\, \epsilon_B \, c_1^2}  \left(  \frac{706\, c_5}{ c_6^{3} \left(\alpha f (\gamma-2) E_l^{\gamma-2} \right)^2\, F_{\rm{peak}}\, D^2}   \right)^{4/2\gamma+13} \nu_{\rm{peak}}^2\, t^2.
\end{equation}

\subsection{Single-epoch spectral modelling}
\label{ssec:single}

In this section we analyse the radio spectrum for individual epochs. The free parameters of the SN--CSM interaction model \citep{10.1086/305676} are the power-law index of the electron energy distribution, $\gamma$; the radius of the forward shock, $R$; and the magnetic field strength, $B$. Theoretical considerations and observations of other  SNe IIP indicate that $\gamma \approx 3.0$ (e.g. \citealt{10.1086/305676} and references within). This implies that the spectral index in the optically thin regime equals $\beta = -1$, although steepening of the spectral slope is possible due to electron cooling. Since the simple SN--CSM interaction model described in \citet{10.1086/305676} does not account for electron cooling, any modelling of a spectral index steeper than $\beta = -1$  when cooling is in effect may incorrectly point to large values of the electron energy distribution power-law index, $\gamma$. Since our analysis suggests that electron cooling is important for SN\,2016X, as we demonstrate in Section~\ref{sec:cooling}, we fix the electron energy distribution power-law index to $\gamma = 3.0$, and fit for the optically thin spectral index, $\beta$, instead. Transforming Equation 1 of \citet{10.1086/305676}, our model reads

\begin{equation}
    \label{eq:beta_chevalier}
    F_{\nu} =  \frac{\pi c_5}{c_6}\, \frac{R^2}{D^2}\, B^{-1/2}\, \left( \frac{\nu}{2 c_1} \right)^{5/2} \left[1 - \exp \left( - \frac{\nu}{\nu_1} \right)^{\beta -5/2} \right],
\end{equation}where $\nu_1$ is the frequency at which the optical depth equals unity for self-absorbed synchrotron:

\begin{equation}
    \label{eq:nu1}
    \nu_1 = 2\, c_1 \left( \frac{c_6}{6 \pi}\, \alpha \, f \, E_l\,  R\,  B^{9/2} \right)^{2/7}.
\end{equation}

\begin{figure}
  \begin{center}
    \includegraphics[width=0.95\linewidth]{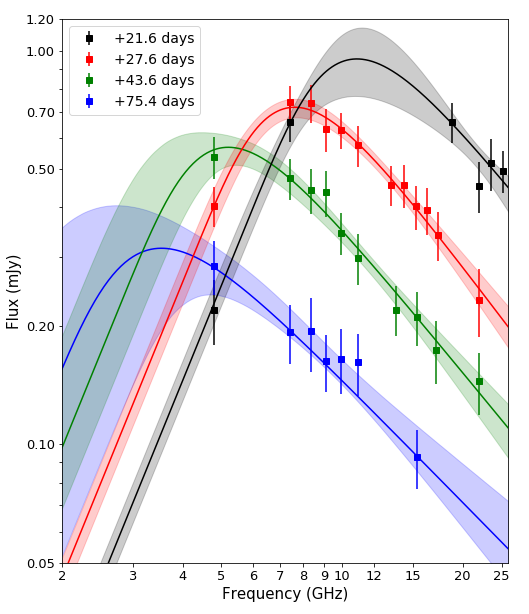}
    \caption{Single-epoch spectral modelling of SN2016X. The shaded regions represent the 1$\sigma$ confidence intervals of the best models.} \label{fig:single}
  \end{center}    
\end{figure}

We implement a Markov chain Monte Carlo (MCMC) algorithm sampling over the optically thin spectral index, $\beta$, the radius of the shock, $R$, and the magnetic field strength, $B$. We determine the best values and the error bars of the free parameters from the posterior probability distributions. We also estimate the confidence intervals of the frequency and flux of the spectral peak from the Markov chains. The velocity of the shock is calculated assuming constant expansion, $ V_{\rm{sh}} = R / t$, and the ratio of the mass-loss rate to the wind velocity is calculated with Equation~\ref{eq:m2w}. The results of the analysis of individual epochs are gathered in Table~\ref{table:single}, and the best-fit models are represented in Figure~\ref{fig:single}.

We note that the optically thick part of the spectrum and the spectral peaks in the later epochs, especially the fourth one, are not well constrained by the modelling, which results in larger error bars in the estimates of the physical quantities. We observe a steepening of the spectral slope $\beta$ in the first three epochs, while the spectral index approaches the expected value $\beta \approx -1$ in the last epoch, which we interpret as a consequence of electron cooling (see Section~\ref{sec:cooling}). The radius of the shock expands from $R \approx 2.6\, \times\,  10^{15}$\,cm at $t=21.6$\,days to $R \approx 5.0\, \times\,  10^{15}$\,cm  at $t=75.4$\,days. The resulting shock velocities, assumed constant in this section ($V_{\rm{sh}} = R/t $), are in  the range  13900 -- 7800\,km\,s$^{-1}$, and are higher than for other SNe IIP observed at ages between 8 and 81 days, which typically show expansion velocities below 10000\, km\,s$^{-1}$ (e.g. \citealt{10.1086/430049, 10.1086/500528, 10.1088/0004-637x/782/1/30}). The high expansion velocities we obtained can indicate that the free expansion phase, where $R \propto t$, does not extend to the time of our observations. Our estimates of the shock radius are compatible with a power law $R \propto t^m$ with index m = 0.8 $\pm$ 0.1, excluding the last epoch. Finally, we obtain a weighted average mass-loss rate of $\dot{M}$\,=\,(4.4 $\pm$ 0.3)\,$\times\, 10^{-7}\, M_{\odot}$\,yr$^{-1}$, for an assumed wind velocity of  $w = 10$\,km\,s$^{-1}$. 

%
\begin{table*}
\caption{Results of the single-epoch spectral analysis.}    
\label{table:single}      
\centering          
\renewcommand{\arraystretch}{1.5}
\begin{tabular}{c c c c c c c c c c }      
\hline\hline       
                     
\multirow{2}{*}{Epoch} & \multirow{2}{*}{$\chi^2$} & \multirow{2}{*}{dof} 
& \multirow{2}{*}{$\beta$}  & \multirow{2}{*}{$\nu_{\rm{peak}}$} & \multirow{2}{*}{$F_{\rm{peak}}$} 
& Magnetic & Shock & Shock & \multirow{2}{*}{$\dot{M} / w$}\\

& & & & & & field & radius & velocity & \\

(days) & & & & (GHz) & (mJy) & (G) & ($10^{15}$\,cm) & ($10^3$\,km\,s$^{-1}$) 
& $  \left( \rm{ 10^{-8} \frac{M_{\odot}\,yr^{-1}}{km\,s^{-1}} } \right) $\\
 
\hline                    

+21.6 & 2.2 & 3 & $-1.26^{+0.40}_{-0.50}$ 
                & $11.0^{+0.4}_{-0.5}$  & $0.93^{+0.22}_{-0.16}$
                & $1.3^{+0.1}_{-0.1}$   & $2.6^{+0.1}_{-0.1}$   
                & $13.9^{-0.6}_{-0.6}$  & $4.9^{+0.4}_{-0.4}$  \\

+27.6 & 1.2 & 9 & $-1.35^{+0.16}_{-0.16}$      
                & $7.6^{+0.3}_{-0.3}$    & $0.72^{+0.04}_{-0.04}$  
                & $1.0^{+0.1}_{-0.1}$    & $3.2^{+0.2}_{-0.2}$ 
                & $13.5^{+0.7}_{-0.7}$   & $4.1^{+0.3}_{-0.3}$   \\

+43.5 & 1.9 & 7 & $-1.25^{+0.19}_{-0.20}$ 
                & $5.3^{+0.5}_{-0.8}$   & $0.57^{+0.06}_{-0.05}$
                & $0.7^{+0.1}_{-0.1}$   & $4.2^{+1.1}_{-0.6}$ 
                & $11.2^{+2.9}_{-1.4}$  & $5.1^{+1.1}_{-1.5}$   \\

+75.4 & 1.9 & 4 & $-1.05^{+0.21}_{-0.25}$ 
                & $3.5^{+1.4}_{-1.0}$   & $0.32^{+0.10}_{-0.06}$ 
                & $0.5^{+0.2}_{-0.1}$   & $5.0^{+3.0}_{-1.8}$ 
                & $7.8^{+4.7}_{-2.9}$   & $7.7^{+7.6}_{-3.9}$   \\

\hline                  
\end{tabular}

{\footnotesize We use a model with $\gamma = 3$ and with free parameters for the optically thin spectral index, $\beta$; the radius of the shock, $R$; and the magnetic field, $B$. The columns in the table (from left to right) are the epoch of the observations; the $\chi^2$; the number of degrees of freedom of the fitting; the spectral slope in the optically thin regime, $\beta$; the frequency and flux of the observed spectral peak, $\nu_{\rm{peak}}$ and $F_{\rm{peak}}$; the magnetic field strength, $B$; the radius of the shock $R$; the shock expansion velocity assuming constant velocity $V_{\rm{sh}}$; and the ratio of the mass-loss rate to the velocity of the wind, $\dot{M}/w$ (Equation~\ref{eq:m2w}). }

\end{table*}


\subsection{Multiple-epoch spectral modelling}
\label{ssec:multiple}

In this section we analyse the radio spectra at all epochs including expressions for the temporal evolution of the radius of the shock and the magnetic field. We assume that the electron energy distribution power-law index ($\gamma = 3$) and the filling factor ($f = 0.5$) remain constant for the entire evolution of the supernova, as observations have shown in other SNe (e.g. \citealt{10.1086/170571}).

We introduce a self-similar solution for the evolution of the radius of shock, $R = R_0\, (t/t_0)^m$, where $m$ is a number between 2/3 and 1 \citep{10.1086/160167}. We also introduce a self-similar solution for the evolution of the magnetic field: $B = B_0\, (t/t_0)^{-b}$. The value of the exponent $b$ is indicative of the amplification mechanism of the magnetic field: $b \approx 1$ points towards turbulent motions in the shocked regions, while $b \approx m$ implies compression of the circumstellar magnetic fields \citep{10.1086/160167}.

\begin{figure}[ht!]
  \begin{center}
    \includegraphics[width=0.9\linewidth]{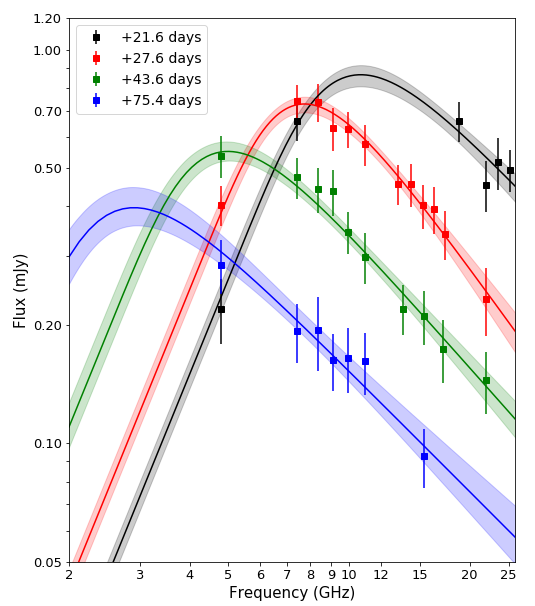}
    \caption{Multiple-epoch spectral modelling of SN2016X. The shaded regions represent the 1$\sigma$ confidence intervals of the best model.} 
    \label{fig:multi}
  \end{center}
\end{figure}

We implement a Markov chain Monte Carlo algorithm, sampling over the spectral slope at individual epochs, $\beta$, the parameters describing the temporal evolution of the radius of the shock, $R_0$ and $m$, and the parameters describing the temporal evolution of the magnetic field, $B_0$ and $b$. We evaluate $R_0$ and $B_0$ at the first epoch of our observations by fixing $t_0$ to $t=21.6$\,days. We estimate the values and the error bars of the free parameters from the posterior probability distributions, for which we obtain $R_0=(2.6 \pm 0.1)\,\times\,10^{15}$\, cm, $m = 0.76^{+0.09}_{-0.07}$, $B_0 = (1.25 \pm 0.06)$\, G, and $b = 1.00 \pm 0.06$ (with a $\chi^2 = 7.8$ and degrees of freedom = 27). We also derive the features of the spectral peaks at individual epochs from the Markov chains. We estimate the magnetic field strength, the shock radius and the expansion velocity from their temporal evolution ($V_{\rm{sh}} = dR/dt = R_0\, m \, t^{m-1}$ for the shock velocity). We calculate the ratio of the mass-loss rate to the wind velocity with a modified version of Equation~\ref{eq:m2w} that includes a factor $m^{-2}$ to account for the evolution of the shock radius. The results of the multiple-epoch analysis are summarised in Table~\ref{table:multi}, and the best-fit model is represented in Figure~\ref{fig:multi}.

The results of the multiple-epoch analysis are similar to those in Section~\ref{ssec:single}, except for the slower expansion velocities in the range 10700 -- 8000\,km\,s$^{-1}$. The temporal evolution of the magnetic field ($b \approx 1$) suggests that turbulent motions are responsible for the amplification mechanism. We obtain a weighted average mass-loss rate of $\dot{M}=$\,(7.8 $\pm$ 0.9)\,$\times\, 10^{-7}\, M_{\odot}$\, yr$^{-1}$, taking into account the error bars and for an assumed wind velocity of $w = 10$\,km\,s$^{-1}$. We also observe spectral steepening during the first three epochs (see Section~\ref{sec:cooling}). Since the spectral peaks are more tightly constrained with the multiple-epoch analysis and the error bars are generally lower, we use the results of this section in Sections~\ref{sec:cooling} and \ref{sec:x-ray}.

%
\begin{table*}
\caption{Results of the multiple-epoch spectral analysis.}    
\label{table:multi}      
\centering          
\renewcommand{\arraystretch}{1.5}
\begin{tabular}{c c c c c c c c }      
\hline\hline       
                     
\multirow{2}{*}{Epoch} & \multirow{2}{*}{$\beta$} & \multirow{2}{*}{$\nu_{\rm{peak}}$} & \multirow{2}{*}{$F_{\rm{peak}}$} & Magnetic & Shock & Shock & \multirow{2}{*}{$\dot{M} / w$}\\

& & & & field & radius & velocity & \\

(days) & & (GHz) & (mJy) & (G) & ($10^{15}$\,cm) & ($10^3$\,km\,s$^{-1}$) 
& $  \left( \rm{ 10^{-8} \frac{M_{\odot}\,yr^{-1}}{km\,s^{-1}} } \right) $\\
 
\hline                    

+21.6 & $-1.08^{+0.15}_{-0.15}$ 
      & $10.7^{+0.4}_{-0.3}$    & $0.86^{+0.05}_{-0.05}$
      & $1.25^{+0.06}_{-0.06}$  & $2.6^{+0.1}_{-0.1}$   
      & $10.7^{+1.1}_{-0.9}$    & $8.1^{+1.6}_{-1.4}$ \\

+27.6 & $-1.42^{+0.14}_{-0.15}$ 
      & $7.8^{+0.2}_{-0.2}$     & $0.73^{+0.03}_{-0.03}$
      & $0.98^{+0.04}_{-0.04}$  & $3.2^{+0.1}_{-0.1}$   
      & $10.1^{+1.2}_{-1.0}$    & $7.2^{+1.6}_{-1.4}$ \\

+43.5 & $-1.15^{+0.10}_{-0.10}$ 
      & $5.0^{+0.2}_{-0.2}$     & $0.55^{+0.03}_{-0.03}$
      & $0.62^{+0.03}_{-0.03}$  & $4.5^{+0.3}_{-0.2}$   
      & $9.1^{+1.4}_{-1.1}$     & $7.9^{+2.1}_{-1.9}$ \\
      
+75.4 & $-1.02^{+0.13}_{-0.13}$ 
      & $2.9^{+0.2}_{-0.2}$     & $0.40^{+0.05}_{-0.04}$
      & $0.36^{+0.03}_{-0.03}$  & $6.7^{+0.7}_{-0.5}$   
      & $8.0^{+1.7}_{-1.2}$     & $8.5^{+2.9}_{-2.4}$ \\
      
\hline                  
\end{tabular}

{\footnotesize We use a model with $\gamma = 3$ and with free parameters for the optically thin spectral indexes, $\beta$; the parameters of the temporal evolution of the shock radius, $R_0$ and $m$ ($R = R_0\,(t/t_0)^{m}$, $t_0$=21.6\,days); and temporal evolution of the magnetic field, $B_0$ and $b$ ($B = B_0\,(t/t_0)^{-b}$, $t_0$=21.6\,days).The columns in the table (from left to right) are the epoch of observation; the optically thin spectral index, $\beta$; the frequency of the spectral peak, $\nu_{\rm{peak}}$; the flux of the spectral peak, $F_{\rm{peak}}$; the magnetic field strength, $B$; the radius of the shock $R$; the expansion velocity, $V_{\rm{sh}}$; and the ratio of the mass-loss rate to the velocity of the wind, $\dot{M}/w$ (Equation~\ref{eq:m2w}).}

\end{table*}


\subsection{Absorption mechanism}
\label{ssec:ffa}

Radio emission from SNe is affected by absorption at early times. Our radio model presented in Section~\ref{sec:analysis} includes self-absorbed synchrotron radiation, but free-free absorption by ionised gas in the CSM can also play an important role. Here we investigate the significance of the free-free absorption for SN\,2016X.

Free-free absorption has been studied in detail (e.g. \citealt{10.1086/160167, 10.1146/annurev.astro.40.060401.093744, 10.1086/507606}), and examined for individual SNe on several occasions (e.g. \citealt{10.1088/0004-637x/746/1/21, 2018ApJ...863..163N}). The optical depth for free-free absorption is calculated as

\begin{equation}
    \label{eq:tau_ffa1}
    \tau_{\rm{ffa}} = 0.083\, T_{\rm{e}}^{-1.35}\,\nu^{-2.1}\,\rm{EM},
\end{equation}where $T_{\rm{e}}$ is the temperature of the electrons in the CSM, the frequency is expressed in GHz, and $\rm{EM}$ is the integral of the square of the electron density along the line of sight expressed in cm$^{-6}$\,pc, known as emission measure. The EM from a radius, $R$, to infinity can be calculated as

\begin{equation}
    \label{eq:em}
    \rm{EM} =  \int_{R}^{\infty} n_0^2 \left( \frac{r}{R_0} \right)^{-4} \,dr =  \frac{1}{3}\,n_0^2\,R_0 \left( \frac{R}{R_0} \right)^{-3},
\end{equation}where $n_0$ is the electron density at radius $R_0$. 

Expressing $n_0$ as a function of $\dot{M}/w$ and substituting the radius by its time evolution solution found in Section~\ref{ssec:multiple} ($R = R_0 (t/t_0)^m$), the free-free optical depth becomes

\begin{equation}
    \label{eq:tau_ffa2}
    \tau_{\rm{ffa}} = 8.1 \times 10^{16}\, \left( \frac{\dot{M}}{w} \right)^{2}\,\nu^{-2.1}\,t^{-2.28},
\end{equation}where we  use $T_e = 10^5$\,K (\citealt{1988A&A...192..221L}, \citealt{10.1086/500528}). For our first epoch, we obtain $\tau_{\rm{ffa}}[\nu = 4.8, 8.4, 22.0\,\rm{GHz}]=[0.018, 0.006, 0.001]$, which shows that the free-free optical depth is not relevant for our observations of SN\,2016X.

\section{Electron cooling}
\label{sec:cooling}

Relativistic electrons can undergo synchrotron or inverse Compton cooling, and affect the spectra and light curves of SNe. In particular inverse Compton cooling can be important for some SNe IIP during the plateau phase due to the  high density of photospheric photons \citep{10.1086/500528}. Our spectral analysis in Section~\ref{sec:analysis} suggests that electron cooling may be significant for SN\,2016X; therefore, we study the characteristic timescales and frequencies of the cooling mechanisms in this section.

First, we compare the timescales of synchrotron and the inverse Compton cooling, $t_{\rm{synch}}$ and $t_{\rm{IC}}$, to the timescale of the expansion of the radio-emitting shell, $t_{\rm{SN}}$. We use Equations 5 and 6 from \citet{10.1086/500528},

\begin{equation}
    t_{\rm{synch}} / t_{\rm{SN}} = 2.0 \times 10^{-6}\, \left( \epsilon_{B}\, \frac{\dot{M}}{w} \right)^{-3/4}\, \left( \frac{t}{\nu} \right)^{1/2} 
,\end{equation}

\begin{equation}
    t_{\rm{IC}} / t_{\rm{SN}} = 3.6\times 10^{35} \, \frac{V_{\rm{sh}}^2}{L_{\rm{bol}}}  \left( \epsilon_{B}\, \frac{\dot{M}}{w} \right)^{1/4}\, \left( \frac{t}{\nu} \right)^{1/2},
\end{equation}where $L_{\rm{bol}}$ is the bolometric luminosity of the supernova in erg\,s$^{-1}$, the velocity of the shock is expressed in km\,s$^{-1}$, the ratio of the mass-loss rate to the wind velocity is expressed in $M_{\odot}$\,yr$^{-1}$/\,km\,s$^{-1}$, times are expressed in days, and frequencies are expressed in GHz. We approximate the bolometric light curve presented in \citet{10.1093/mnras/sty066} by a power law, $L_{\rm{bol}}(t) \propto t^\delta$, with $\delta = -0.59$, between phases $t$\,=\,30\,days ($L_{\rm{bol}}(30) = 6.3 \times 10^{41}$\, erg\, s$^{-1}$)  and $t$\,=\,80\,days ($L_{\rm{bol}}(80) = 3.5\, \times\, 10^{41}$\, erg\, s$^{-1}$). Assuming $\epsilon_{B} = 0.1$, we obtain $t_{\rm{synch}} / t_{\rm{SN}} \approx$ 2.4 -- 9.5 and $t_{\rm{IC}} / t_{\rm{SN}} \approx$ 0.4 -- 1.7 for the frequencies and the times of our observations ($\nu$\,=\,4.8 -- 22.0\,GHz, $t$\,=\, 21.6 --75.4\, days). We note that the shorter cooling timescales are obtained at earlier times and higher frequencies. These timescales indicate that inverse Compton cooling can be important for SN\,2016X, which in turn may indicate low values of the mass-loss rate, $\dot{M}$, and $\epsilon_B$ \citep{10.1086/500528}.

Second, we estimate the cooling frequencies for both mechanisms. The synchrotron cooling frequency is

\begin{equation}
    \nu_{\rm{synch}} = 1.8 \times 10^{-8}\,\pi\, \frac{c\, m_e\, q_e}{\sigma_T^2}\, B^{-3}\, t^{-2}\quad \rm{GHz},
\end{equation}where $m_e$ and $q_e$ are the mass and charge of the electron, $c$ is the speed of light, $\sigma_T$ is the Thomson cross-section, and all variables are expressed in cgs units.
The inverse Compton cooling frequency is

\begin{equation}
    \nu_{\rm{IC}} = 1.3\times 10^{71} \, \frac{V_{\rm{sh}}^4}{L_{\rm{bol}}^2}  \left( \epsilon_{B}\, \frac{\dot{M}}{w} \right)^{1/2}\quad \rm{GHz}
,\end{equation}where the bolometric luminosity is expressed in erg\,s$^{-1}$, the velocity of the shock is expressed in km\,s$^{-1}$, and the ratio of the mass-loss rate to the wind velocity is expressed in $M_{\odot}$\,yr$^{-1}$/\,km\,s$^{-1}$.

We obtain characteristic synchrotron cooling frequencies above $\nu_{\rm{synch}} \gtrsim$ 200\, GHz for all our epochs, so our observations are not affected by synchrotron cooling. We obtain inverse Compton cooling frequencies in the range $\nu_{\rm{IC}}$ = 2.8 -- 14.7 \,GHz, which can affect our observations. We interpret the steepening of the optically thin spectral index at the earlier epochs, especially at $t = 27.6$\,days for which $\beta \approx -1.4$, as the signature of inverse Compton cooling in the radio spectrum.

\section{X-ray emission and energy equipartition}
\label{sec:x-ray}

In this section we study the origin of the X-ray emission from SN\,2016X, and we combine the X-ray data with our radio analysis results to verify the energy equipartition assumption. We use the X-ray luminosity registered by \textit{Chandra},  $L_X = 4.7^{+1.4}_{-1.7}\,\times\,10^{38}$\,erg\,s$^{-1}$ in the 0.3--10\,keV band at phase $t = 18.9$\, days \citep{10.3847/2041-8213/ab0558}, and we assume that the X-ray luminosity remains constant until the time of our first radio observations at $t=21.6$\, days. 

First, we consider thermal emission from the circumstellar and reverse shocks as the source of the X-ray emission. We estimate the temperatures behind the circumstellar shock, $T_{\rm{cs}}$, and the reverse shock, $T_{\rm{rs}}$, as \citep{1996ApJ...461..993F}

\begin{equation}
    T_{\rm{cs}} = 22.7 \, m^2 \, V_{\rm{sh}}^2 = 1.5^{+0.7}_{-0.5} \times 10^9\,\rm{K} \qquad [m=0.76]
,\end{equation}
\begin{equation}
    T_{\rm{rs}} = \left( 1 - \frac{1}{m} \right)^2\, T_{\rm{cs}} = 1.5^{+3.0}_{-1.1} \times 10^8\,\rm{K} \quad \; [m=0.76]
,\end{equation}where $m$ is the exponent of the evolution of the shock radius ($R \propto t^m$), and $V_{\rm{sh}}$ is expressed in km\,s$^{-1}$.

With these temperatures, we calculate the X-ray emission per unit energy (Equations 8 and 9 from \citealt{10.1086/500528}),

\begin{equation}
    \frac{dL_{\rm{cs}}}{dE} = 4.0 \times 10^{47}\, \frac{E_X^{-0.23}\, T_{\rm{cs}}^{-0.08}}{\exp{(0.0116\, E_X / T_{\rm{cs}})}} \, \left( \dfrac{\dot{M}}{w} \right)^2 V_{\rm{sh}}^{-1}\, t^{-1} 
,\end{equation}

\begin{equation}
    \frac{dL_{\rm{rs}}}{dE} = 5.9 \times 10^{48}\, \frac{T_{\rm{rs}}^{-0.024}}{\exp{(0.0116 / T_{\rm{rs}}})} \, \left( \dfrac{\dot{M}}{w} \right)^2 V_{\rm{sh}}^{-1}\, t^{-1}
\end{equation}in erg\,s$^{-1}$\,keV$^{-1}$ units. Here $E_X = 4.9$\,keV is the average energy of the X-ray photon, the temperatures are expressed in $10^9$\,K units, the ratio of the mass-loss rate to the wind velocity is expressed in $M_{\odot}$\,yr$^{-1}$/\,km\,s$^{-1}$, the velocity of the shock is expressed in km\,s$^{-1}$, and the age of the SN is expressed in days. Adding these two contributions, we obtain a thermal X-ray emission of $\sim 6.8 \times 10^{35}$\,erg\,s$^{-1}$\,keV$^{-1}$, for which the uncertainty is dominated by the estimates of the temperatures. This can roughly account for 1\% of the X-ray emission observed with {\it Chandra}. 

X-ray thermal emission can be absorbed by a dense shell of cool low-ionised plasma in the contact discontinuity region if the reverse shock is radiative \citep{1996ApJ...461..993F}. Radiative reverse shocks are plausible when the supernova presents large density gradients (i.e. $n \gtrsim 9 $ with $\rho_{\rm{SN}} \propto R^{-n}$), but we find $n = (2m-3)/(m-1) \approx 6.2$ for SN\,2016X. Moreover, the presence of a layer of cool absorbing plasma also requires short cooling times for free-free emission behind the reverse shock; however, for SN\,2016X we find

\begin{equation}
    t_{\rm{cool}} = 5 \times 10^{-19}\, \left( \dfrac{\dot{M}}{w} \right)^{-1} V_{\rm{sh}}^{3}\, t^{2} = 3650\, \rm{days}
.\end{equation}
We therefore conclude that absorption of X-ray emission from the reverse shock is not relevant for SN\,2016X.

Since the thermal emission cannot explain the observed X-ray flux alone, we consider emission via inverse Compton scattering of photospheric photons. The X-ray emission per unit energy due to inverse Compton scattering is (Equation 11 from \citealt{10.1086/500528})

\begin{equation}
    \label{eq:xic}
    \frac{dL_{\rm{IC}}}{dE} = 8.8\, \Gamma_{\rm{min}}\, \epsilon_e\, E_X^{-1}\, L_{\rm{bol}}\, \frac{\dot{M}}{w}\, V_{\rm{sh}}\, t^{-1} \quad \rm{erg}\,\rm{s}^{-1}\,\rm{keV}^{-1}
,\end{equation}where $\Gamma_{\rm{min}}$ is the minimum Lorentz factor of the electron population, and $L_{\rm{bol}} = 10^{42}$\,erg\,s$^{-1}$ is the bolometric luminosity at $t = 21.6$\,days \citep{10.1093/mnras/sty066}. We assume that the minimum Lorentz factor is $\Gamma_{\rm{min}}=1$ as the population of non-relativistic electrons can be significant for all subtypes of SNe II \citep{10.1086/507606}. We obtain an X-ray luminosity via inverse Compton scattering of $6.2 \times 10^{36}$\,erg\,s$^{-1}$\,keV$^{-1}$, which accounts for approximately 14\% of the observed emission with {\it Chandra}. We note that the X-ray luminosity could have decreased at the epoch of our first radio observations. We also note that these calculations are sensitive to the energy equipartition assumption as a deviation from equipartition leads to an underestimation of the mass-loss rate divided by the wind velocity.


Hence, we investigate potential deviations from energy equipartition combining the X-ray and radio observations. From Equations~\ref{eq:b} and \ref{eq:b_scaling} it can be seen that

\begin{equation}
    S_{*} \equiv \frac{\dot{M}}{w} \,\epsilon_B\, \alpha^{8/19}
    = 2.4\times 10^{18}\, \left( \frac{\nu_{\rm{peak}}\, t}{ m} \right)^2\, \left(F_{\rm{peak}}\, D^2 \right)^{-4/19} 
.\end{equation}Using the results of Section~\ref{ssec:multiple}, we obtain $S_{*} = 5.2 \times 10^{12}$\,g\,cm$^{-1}$. Assuming that inverse Compton cooling is responsible for the observed X-ray flux, we can rewrite Equation~\ref{eq:xic} substituting the mass-loss rate term by $S_{*}$,

\begin{equation}
    E_X\, \frac{dL_{\rm{L_X}}}{dE} = 8.8\, \Gamma_{\rm{min}}\, S_{*}\, \alpha^{11/19}\, L_{\rm{bol}}\, V_{\rm{sh}}\, t^{-1}
,\end{equation} from which we can obtain the ratio $\alpha = \epsilon_e/\epsilon_B$. For $\Gamma_{\rm{min}} =1$,

\begin{equation}
    \alpha = \left( 0.114\,E_X\, \frac{dL_{\rm{X}}}{dE}\, \frac{t}{S_{*}\, V_{\rm{sh}} \, L_{\rm{bol}}} \right)^{19/11},
\end{equation}we obtain $\alpha = 28.4$, and assuming $\epsilon_e = 0.1$, we get $\epsilon_B = 0.0035 $. This result represents a moderate departure from the energy equipartition assumption compared to other SNe where $\alpha \gtrsim 100,$ such as SN\,2011dh \citep{2012ApJ...752...78S, 10.1093/mnras/stt1645}, SN\,2012aw \citep{10.1088/0004-637x/782/1/30}, or SN\,2020oi \citep{Horesh_2020}. This deviation from the energy equipartition assumption implies that our radio analysis overestimated the shock radii by $\sim$19\%, and the magnetic field strengths by a factor of $\sim$2, and underestimated the ratio of the mass-loss rate to the wind velocity by a factor of $\sim$6.9.

\section{Discussion}
\label{sec:discussion}

\subsection{Radio analysis}
\label{ssec:disc_radio}

We presented radio observations of the type IIP supernova SN\,2016X obtained with the VLA. We performed a spectral analysis of these observations using the SN--CSM interaction model and modelling the emission as self-absorbed synchrotron radiation. We modelled both individual epochs and multiple epochs introducing self-similar solutions for the evolution of the radius of the shock and the magnetic field. The results of the two analyses are compatible with each other, although the multiple-epoch analysis better constrains  the spectral peaks, which is expected given the limited number of data points for individual epochs.

The expansion properties of SN\,2016X are comparable to other  SNe IIP. We calculated from the multiple-epoch analysis that the radius of the shock expands from (2.6 $\pm$ 0.1)\,$\times 10^{15}$\,cm to (6.7 $\pm$ 0.7)\,$\times 10^{15}$\,cm during the epochs of our observations, evolving according to a power law,  $R \propto t^m$, with index $m = 0.76^{+0.09}_{-0.07}$. The resulting expansion velocities are between 10700\,$\pm$\,1100\,km\,s$^{-1}$ and 8000\,$\pm$\,1700\,km\,s$^{-1}$. The expansion velocities are slightly higher than for other SNe IIP observed at the same ages (e.g. \citealt{10.1086/500528}), although this result depends on the equipartition assumption. The observed evolution of the magnetic field can be explained by the turbulent motion of particles in the shocked regions. 

Assuming a wind velocity of $w = 10$\,km\,s$^{-1}$ and a self-similar temporal evolution of the shock, we obtained a mass-loss rate of $\dot{M}=$\,(7.8 $\pm$\, 0.9)\,$\times\, 10^{-7}\, \alpha^{-8/19}\, (\epsilon_B/0.1)^{-1} M_{\odot}$\, yr$^{-1}$. This mass-loss rate is compatible with the expectations of RSGs and comparable to other SNe IIP, such as  SN\,1999gi \citep{Schlegel_2001}, SN\,2004dj \citep{2018ApJ...863..163N}, or SN\,2011ja \citep{10.1088/0004-637x/774/1/30}, if equipartition is assumed. 

We assessed the significance of free-free absorption by estimating the free-free optical depth at our first epoch of observations. For our lowest frequency, the free-free optical depth is $\tau_{\rm{ffa}}(\nu=4.8\,\rm{GHz})$\,=\,0.018. Hence, we conclude that self-absorbed synchrotron is the dominant absorption mechanism. 

\citet{10.3847/2041-8213/ab0558} interpreted double-peaked Balmer lines in the  SN\,2016X spectrum as a result of bipolar ejecta or asymmetric wind distribution. Our radio analysis incorporates possible deviations from spherical symmetries through the volume filling factor, $f$. However, the flat functional dependence of the physical parameters on the filling factor, $R \propto f^{-1/19}$ and $B \propto f^{-4/19}$, does not allow for an accurate determination. For instance, reducing the filling factor by half only changes the result of the shock radius by less than 4\%. Since the CSM density for SN\,2016X is proportional to $R^{-2.1}$, in agreement with the density profile of spherically symmetric wind ($\propto R^{-2}$), we see no evidence of asymmetric plasma distribution in our radio data. It should also be borne in mind that the only evidence that supports the claim of asphericity (i.e. the double-peaked optical lines) only appears when the nebular phase is well established ($t > 200$\,days), which may not be relevant for our radio observations.

\subsection{Electron cooling and equipartition}
\label{ssec:disc_equipartition}

The characteristic cooling timescales and frequencies suggest that inverse Compton electron cooling is important for SN\,2016X. The signature of inverse Compton cooling is evident in the radio spectrum, where we observe a steepening of the optically thin spectral slope at early times, when electron cooling is most important. This behaviour has been observed in other SNe of different types (e.g. SN\,2002ap, \citealt{2004ApJ...605..823B}; SN\,2012aw, \citealt{10.1088/0004-637x/782/1/30}; SN\,2020oi, \citealt{Horesh_2020}).

Combining our results from the radio analysis with X-ray observations reported by \citet{10.3847/2041-8213/ab0558}, we find a moderate deviation from energy equipartition by a factor $\alpha = 28.4$. Assuming that the fraction of the energy invested in particle acceleration is $\epsilon_e = 0.1$, the fraction of the post-shock energy deposited in magnetic field intensity is $\epsilon_B = 0.0035$. This deviation is modest compared to other SNe, where $\alpha \gtrsim 100$ (e.g. SN\,2011dh \citealt{2012ApJ...752...78S, 10.1093/mnras/stt1645}; SN\,2012aw \citealt{10.1088/0004-637x/782/1/30}; SN\,2020oi \citealt{Horesh_2020}). This deviation from equipartition implies that our results overestimate the shock radii and magnetic field and underestimate the mass-loss rate. The corrected mass-loss rate is $\dot{M}=$\,(5.4 $\pm$\, 0.6)\,$\times\, 10^{-6}\, M_{\odot}$\, yr$^{-1}$, which is similar to that of SN\,1999em \citep{2002ApJ...572..932P}, SN\,2002hh \citep{2006A&A...449..171N}, or SN\,2004et \citep{2007MNRAS.381..280M}. 

\begin{figure}[ht!]
  \begin{center}
    \includegraphics[width=1.0\linewidth]{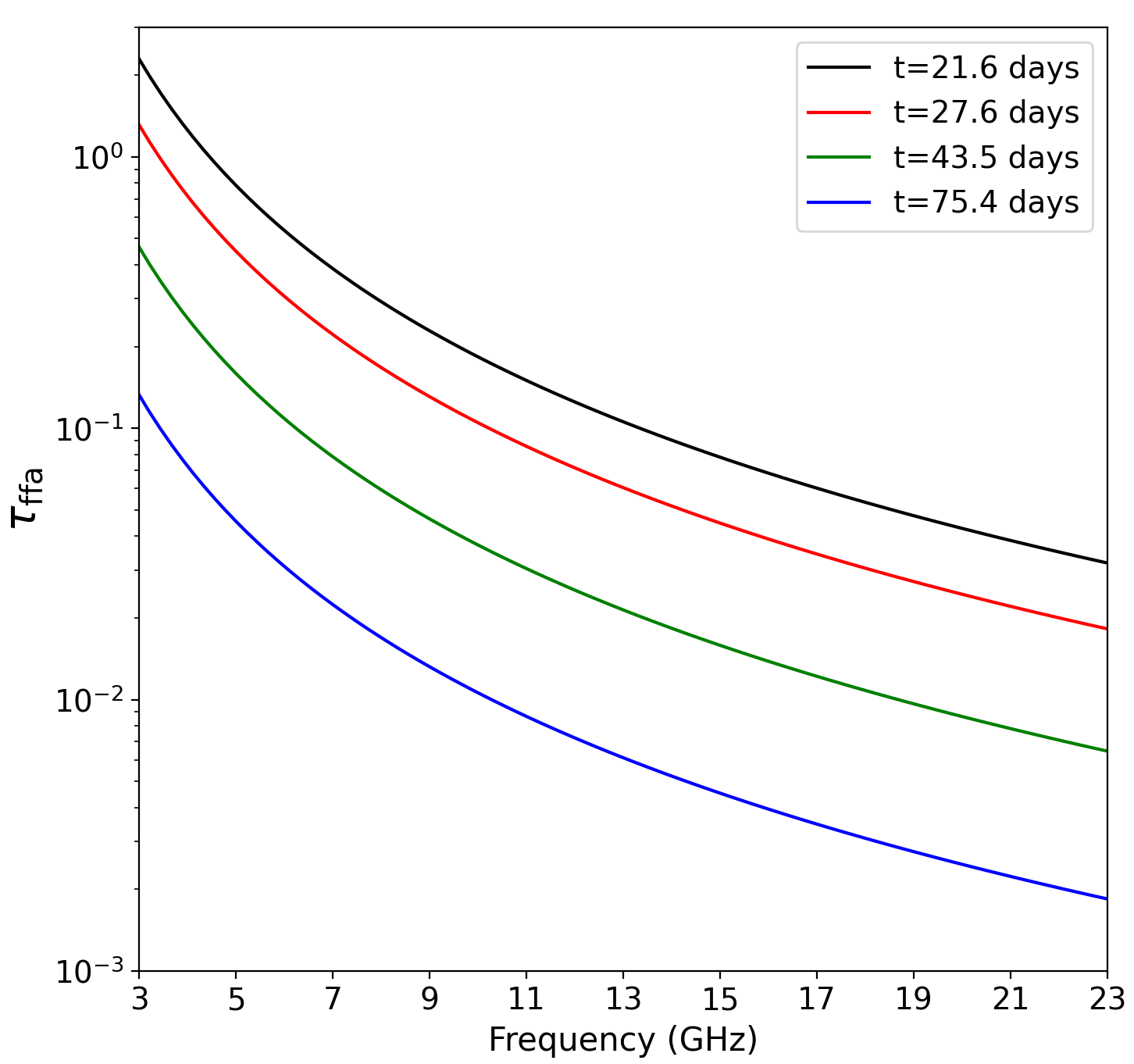}
    \caption{Free-free optical depth as a function of frequency for the epochs of observation of SN\,2016X.}
    \label{fig:tau_ffa}
  \end{center}
\end{figure}

Since the free-free optical depth depends on the mass-loss rate, we review our calculations here. In Figure~\ref{fig:tau_ffa} we present the optical depth as a function of frequency for our epochs of observations. The maximum value of the free-free optical depth is $\tau_{\rm{ffa}}(\nu = 4.8\,\rm{GHz}, t = 21.6\,\rm{d}) = 0.85$, which is below unity. Additionally, we can check if free-free absorption is important by verifying whether the shell velocities obtained from the radio model are lower than the velocities derived from optical spectral lines. For SN\,2016X, we obtained shock velocities of 10700 -- 8000\,km\,s$^{-1}$, while \citet{10.1093/mnras/sty066} inferred velocities in the range $\sim$9000 -- 6000\,km\,s$^{-1}$ between $t$=20 and $t$=80 days from H$\alpha$ lines. With these results, we argue that the self-absorbed synchrotron is the dominant absorption mechanism.

\subsection{SN\,2016X compared to other SNe IIP}
\label{ssec:disc_IIPs}

SN\,2016X is one of the few SNe IIP that are detected in both radio and X-rays, and it is arguably one of the best monitored in radio bands since we collected 36 data points in several frequencies and four different epochs. The other SNe IIP with similar or better radio coverage are SN\,2004dj \citep{2018ApJ...863..163N}, SN\,2004et \citep{2007MNRAS.381..280M}, and SN\,2012aw \citep{10.1088/0004-637x/782/1/30}.

\begin{figure}[ht!]
  \begin{center}
    \includegraphics[width=0.95\linewidth]{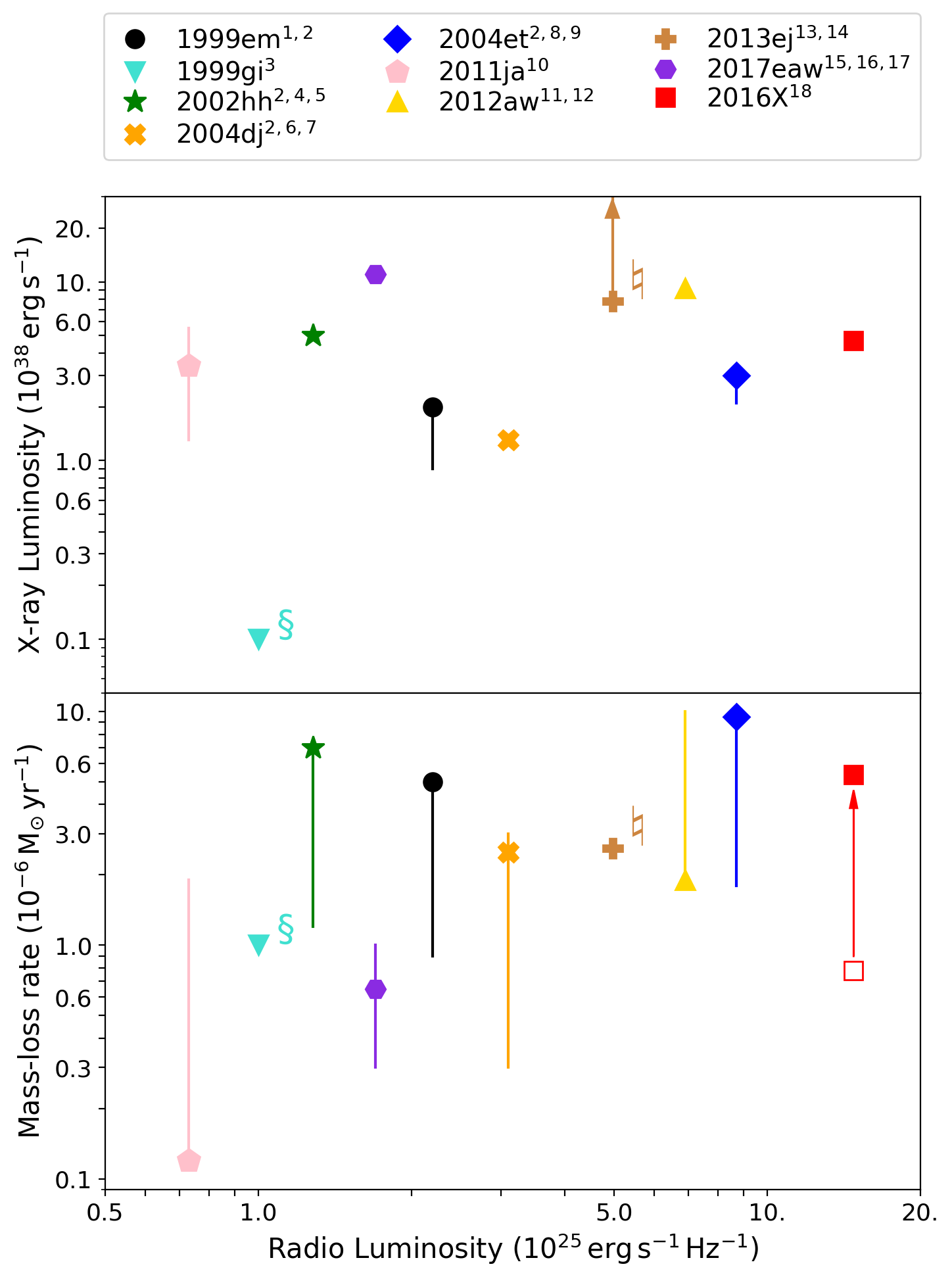}
    \caption{Radio and X-ray properties of SNe IIP. The radio luminosity at $\sim$5\,GHz$^{\natural}$ is measured at the peak of the light curve for SN\,1999em, SN\,2004dj, SN\,2004et, SN\,2012aw, and SN\,2016X; and limits on the flux peak are used for the rest of the  SNe$^{\S}$. The X-ray luminosity is taken as close to the peak as possible in the energy ranges   $\sim$0.4--8\,keV for SN\,1999em, SN\,2002hh, SN\,2004dj, SN\,2004et, SN\,2013ej, and $\sim$0.3--10\,keV for the rest of the  SNe. The mass-loss rate is estimated from radio observations, except for SN\,1999gi, SN\,2013ej, and SN\,2017eaw. For SN\,2016X we plot the mass-loss rate assuming equipartition (empty symbol) and  departure from equipartition (filled symbol). The vertical lines are not error bars, but rather ranges of possible values reported by different authors (e.g. \citealt{2018ApJ...863..163N}).\hspace{\textwidth} \textit{References: } [1] \citealt{2002ApJ...572..932P}, [2] \citealt{10.1086/500528}, [3] \citealt{2001ApJ...556L..25S}, [4] \citealt{2002IAUC.8024....2P}, [5] \citealt{2005IAUC.8572....1B}, [6] \citealt{2005ApJ...623L..21B}, [7] \citealt{2012ApJ...761..100C}, [8] \citealt{2004IAUC.8435....3B}, [9] \citealt{2007MNRAS.381..280M}, [10] \citealt{2013ApJ...774...30C}, [11] \citealt{2016ApJ...817...22C}, [12] \citealt{2021ApJ...908...75B}, [13] \citealt{2012ATel.3995....1I}, [14] \citealt{10.1088/0004-637x/782/1/30}, [15] \citealt{2017ATel10472....1A}, [16] \citealt{2017ATel10427....1G}, [17] \citealt{2019ApJ...876...19S},  [18] This work. \hspace{\textwidth} \textit{Notes: } $^{\natural}$ SN\,2013ej was observed in radio at 9\,GHz. $^{\S}$ SN\,1999gi was not detected in radio, so it is plotted at the arbitrary radio luminosity $L_{\rm{radio}}= 10^{25}$\,erg\,s$^{-1}$\,Hz$^{-1}$.}
    \label{fig:context}
  \end{center}
\end{figure}

In this section we compare the properties of SN\,2016X to other SNe IIP observed in radio and X-ray bands. We present an overview of these SNe IIP in our Figure~\ref{fig:context}, updating the sample in \citet{2018ApJ...863..163N}'s Figure 5. SN\,2016X is the most luminous SN IIP in the radio, with $L_{4.8\,\rm{GHz}} \approx 1.5 \times 10^{26}$\,erg\,s$^{-1}$\,Hz$^{-1}$ ($t=43.5$\,d). The X-ray luminosity of SN\,2016X is average, larger than for SN\,1999gi, SN\,1999em, SN\,2004dj, or SN\,2004et, similar to SN\,2011ja and SN\,2002hh, and smaller than for SN\,2012aw, SN\,2017eaw, and SN\,2013ja. While the range of radio luminosities is restricted to roughly one order of magnitude, the values of X-ray luminosity span   four orders of magnitude.

The mass-loss rate of SN IIP is well constrained within $\sim$1--10\,$\times\, 10^{-6}\, M_{\odot}$\, yr$^{-1}$ for an assumed wind velocity of 10\,km\,s$^{-1}$. SN\,2016X exhibits a mass-loss rate towards the lower limit if equipartition is assumed, but comparable to the highest mass-loss rates taking into account deviations from equipartition. 

The radio spectrum of SN\,2016X is also affected by electron cooling via inverse Compton scattering, such as SN\,2004dj, SN\,2004et, and SN\,2012aw, which may suggest that it is a common process for SNe IIP. SN\,2016X also has  in common with SN\,2012aw the fact that both exhibit deviations from energy equipartition, although the deviation is more severe for SN\,2012aw ($\alpha \sim 112$). The mass-loss rate of SN\,2016X is higher than that of SN\,2012aw by a factor of $\sim$3. The case of SN\,2016X illustrates the importance of obtaining early radio and X-ray observations of young SNe.

\section{Summary}
\label{sec:summary}

In this paper we presented radio observations at frequencies 4.8 -- 25.0\,GHz obtained at four different epochs at ages in the range 21.6--75.4\,days, which makes it one of the best monitored SN IIP  in radio frequencies. The observations can be well modelled with a self-absorbed synchrotron model, introducing self-similar expressions for the time evolution of the shock radius and magnetic field. The results are consistent with the interaction of ejecta with density profile $\rho_{\rm{SN}}\propto R^{-6.2}$ with a CSM with density profile $\rho_{\rm{w}}\propto R^{-2.1}$. The expansion of the shock radius shows moderate deceleration, $R\propto t^{-0.76}$.

We observed steeping of the spectral index in the optically thin regime at earlier epochs, which we interpreted as the signature of electron cooling via inverse Compton scattering. We combined our radio data with X-ray observations, and we found a mild deviation from energy equipartition, with $\alpha = \epsilon_e / \epsilon_B \approx 28.4$. This is the second SN IIP to exhibit this phenomenology, after SN\,2012aw.

\begin{acknowledgements}
 
We thank the referee for a thorough assessment of our work and for helping improve this paper.

RRC acknowledges support from the Lady Davis Trust Fellowship and the Emily Erskine Endowment Funds at the Hebrew University of Jerusalem.

A.H. is grateful for the support by grants from the Israel Science Foundation, the US-Israel Binational Science Foundation (BSF), and the I-CORE Program of the Planning and Budgeting Committee and the Israel Science Foundation. 

The VLA is a component of the National Radio Astronomy Observatory, a facility of the National Science Foundation operated under cooperative agreement by Associated Universities, Inc.

We have made extensive use of the data reduction software {\sc casa} \citep{2007ASPC..376..127M}, and the {\sc python} packages:
{\sc astropy} \citep{2013A&A...558A..33A},  
{\sc numpy} \citep{Oliphant:2006wm}, 
{\sc matplotlib} \citep{Hunter:2007},
{\sc emcee} \citep{2013PASP..125..306F}, and
{\sc corner} \citep{corner}.
        
\end{acknowledgements}

%
%


\bibliographystyle{aa}
\bibliography{BIB/SN2016X_refs}

\begin{thebibliography}{85}
\expandafter\ifx\csname natexlab\endcsname\relax\def\natexlab#1{#1}\fi

\bibitem[{{Argo} {et~al.}(2017){Argo}, {Torres}, {Beswick}, \&
  {Wrigley}}]{2017ATel10472....1A}
{Argo}, M., {Torres}, M.~P., {Beswick}, R., \& {Wrigley}, N. 2017, The
  Astronomer's Telegram, 10472, 1

\bibitem[{{Arnett}(1980)}]{1980ApJ...237..541A}
{Arnett}, W.~D. 1980, \apj, 237, 541

\bibitem[{{Astropy Collaboration} {et~al.}(2013){Astropy Collaboration},
  {Robitaille}, {Tollerud}, {Greenfield}, {Droettboom}, {Bray}, {Aldcroft},
  {Davis}, {Ginsburg}, {Price-Whelan}, {Kerzendorf}, {Conley}, {Crighton},
  {Barbary}, {Muna}, {Ferguson}, {Grollier}, {Parikh}, {Nair}, {Unther},
  {Deil}, {Woillez}, {Conseil}, {Kramer}, {Turner}, {Singer}, {Fox}, {Weaver},
  {Zabalza}, {Edwards}, {Azalee Bostroem}, {Burke}, {Casey}, {Crawford},
  {Dencheva}, {Ely}, {Jenness}, {Labrie}, {Lim}, {Pierfederici}, {Pontzen},
  {Ptak}, {Refsdal}, {Servillat}, \& {Streicher}}]{2013A&A...558A..33A}
{Astropy Collaboration}, {Robitaille}, T.~P., {Tollerud}, E.~J., {et~al.} 2013,
  \aap, 558, A33

\bibitem[{{Beswick} {et~al.}(2005){Beswick}, {Fenech}, {Thrall}, {Argo},
  {Muxlow}, \& {Pedlar}}]{2005IAUC.8572....1B}
{Beswick}, R.~J., {Fenech}, D., {Thrall}, H., {et~al.} 2005, \iaucirc, 8572, 1

\bibitem[{{Beswick} {et~al.}(2004){Beswick}, {Muxlow}, {Argo}, {Pedlar}, \&
  {Marcaide}}]{2004IAUC.8435....3B}
{Beswick}, R.~J., {Muxlow}, T.~W.~B., {Argo}, M.~K., {Pedlar}, A., \&
  {Marcaide}, J.~M. 2004, \iaucirc, 8435, 3

\bibitem[{Beswick {et~al.}(2005)Beswick, Muxlow, Argo, Pedlar, Marcaide, \&
  Wills}]{10.1086/430049}
Beswick, R.~J., Muxlow, T. W.~B., Argo, M.~K., {et~al.} 2005, The Astrophysical
  Journal, 623, L21

\bibitem[{{Beswick} {et~al.}(2005){Beswick}, {Muxlow}, {Argo}, {Pedlar},
  {Marcaide}, \& {Wills}}]{2005ApJ...623L..21B}
{Beswick}, R.~J., {Muxlow}, T.~W.~B., {Argo}, M.~K., {et~al.} 2005, \apjl, 623,
  L21

\bibitem[{Bietenholz {et~al.}(2021)Bietenholz, Bartel, Argo, Dua, Ryder, \&
  Soderberg}]{10.3847/1538-4357/abccd9}
Bietenholz, M.~F., Bartel, N., Argo, M., {et~al.} 2021, The Astrophysical
  Journal, 908, 75

\bibitem[{{Bietenholz} {et~al.}(2021){Bietenholz}, {Bartel}, {Argo}, {Dua},
  {Ryder}, \& {Soderberg}}]{2021ApJ...908...75B}
{Bietenholz}, M.~F., {Bartel}, N., {Argo}, M., {et~al.} 2021, \apj, 908, 75

\bibitem[{{Bj{\"o}rnsson} \& {Fransson}(2004)}]{2004ApJ...605..823B}
{Bj{\"o}rnsson}, C.-I. \& {Fransson}, C. 2004, \apj, 605, 823

\bibitem[{{Bock} {et~al.}(2016){Bock}, {Shappee}, {Stanek}, {Prieto},
  {Kochanek}, {Holoien}, {Brown}, {Godoy-Rivera}, {Basu}, {Bersier}, {Dong},
  {Chen}, {Brimacombe}, {Masi}, \& {Kiyota}}]{2016ATel.8566....1B}
{Bock}, G., {Shappee}, B.~J., {Stanek}, K.~Z., {et~al.} 2016, The Astronomer's
  Telegram, 8566, 1

\bibitem[{Bose {et~al.}(2019)Bose, Dong, Elias-Rosa, Shappee, Bersier, Benetti,
  Stritzinger, Grupe, Kochanek, Prieto, Chen, Kuncarayakti, Mattila,
  Morales-Garoffolo, Morrell, Onori, Reynolds, Siviero, Somero, Stanek,
  Terreran, Thompson, Tomasella, Ashall, Gall, Gromadzki, \&
  Holoien}]{10.3847/2041-8213/ab0558}
Bose, S., Dong, S., Elias-Rosa, N., {et~al.} 2019, The Astrophysical Journal,
  873, L3

\bibitem[{Bruch {et~al.}(2020)Bruch, Gal-Yam, Schulze, Yaron, Yang, Soumagnac,
  Rigault, Strotjohann, Ofek, Sollerman, Masci, Barbarino, Ho, Fremling,
  Perley, Nordin, Cenko, Adams, Adreoni, Bellm, Blagorodnova, Bulla, Burdge,
  De, Dhawan, Drake, Duev, Dugas, Graham, Graham, Jencson, Karamehmetoglu,
  Kasliwal, Kim, Kulkarni, Kupfer, Mahabal, Miller, Prince, Riddle, Sharma,
  Smith, Taddia, Taggart, Walters, \& Yan}]{Bruch20}
Bruch, R.~J., Gal-Yam, A., Schulze, S., {et~al.} 2020, arXiv
  [\eprint{2008.09986}]

\bibitem[{{Chakraborti} {et~al.}(2016){Chakraborti}, {Ray}, {Smith},
  {Margutti}, {Pooley}, {Bose}, {Sutaria}, {Chandra}, {Dwarkadas}, {Ryder}, \&
  {Maeda}}]{2016ApJ...817...22C}
{Chakraborti}, S., {Ray}, A., {Smith}, R., {et~al.} 2016, \apj, 817, 22

\bibitem[{{Chakraborti} {et~al.}(2013){Chakraborti}, {Ray}, {Smith}, {Ryder},
  {Yadav}, {Sutaria}, {Dwarkadas}, {Chandra}, {Pooley}, \&
  {Roy}}]{2013ApJ...774...30C}
{Chakraborti}, S., {Ray}, A., {Smith}, R., {et~al.} 2013, \apj, 774, 30

\bibitem[{Chakraborti {et~al.}(2013)Chakraborti, Ray, Smith, Ryder, Yadav,
  Sutaria, Dwarkadas, Chandra, Pooley, \& Roy}]{10.1088/0004-637x/774/1/30}
Chakraborti, S., Ray, A., Smith, R., {et~al.} 2013, The Astrophysical Journal,
  774, 30

\bibitem[{{Chakraborti} {et~al.}(2012){Chakraborti}, {Yadav}, {Ray}, {Smith},
  {Chandra}, \& {Pooley}}]{2012ApJ...761..100C}
{Chakraborti}, S., {Yadav}, N., {Ray}, A., {et~al.} 2012, \apj, 761, 100

\bibitem[{{Chevalier}(1976)}]{1976ApJ...207..872C}
{Chevalier}, R.~A. 1976, \apj, 207, 872

\bibitem[{Chevalier(1981)}]{10.1086/159460}
Chevalier, R.~A. 1981, The Astrophysical Journal, 251, 259

\bibitem[{Chevalier(1982)}]{10.1086/160167}
Chevalier, R.~A. 1982, The Astrophysical Journal, 259, 302

\bibitem[{Chevalier(1998)}]{10.1086/305676}
Chevalier, R.~A. 1998, The Astrophysical Journal, 499, 810

\bibitem[{Chevalier \& Fransson(2006)}]{10.1086/507606}
Chevalier, R.~A. \& Fransson, C. 2006, The Astrophysical Journal, 651, 381

\bibitem[{Chevalier {et~al.}(2006)Chevalier, Fransson, \&
  Nymark}]{10.1086/500528}
Chevalier, R.~A., Fransson, C., \& Nymark, T.~K. 2006, The Astrophysical
  Journal, 641, 1029

\bibitem[{{Corsi} {et~al.}(2014){Corsi}, {Ofek}, {Gal-Yam}, {Frail},
  {Kulkarni}, {Fox}, {Kasliwal}, {Sullivan}, {Horesh}, {Carpenter}, {Maguire},
  {Arcavi}, {Cenko}, {Cao}, {Mooley}, {Pan}, {Sesar}, {Sternberg}, {Xu},
  {Bersier}, {James}, {Bloom}, \& {Nugent}}]{2014ApJ...782...42C}
{Corsi}, A., {Ofek}, E.~O., {Gal-Yam}, A., {et~al.} 2014, \apj, 782, 42

\bibitem[{Foley {et~al.}(2007)Foley, Smith, Ganeshalingam, Li, Chornock, \&
  Filippenko}]{10.1086/513145}
Foley, R.~J., Smith, N., Ganeshalingam, M., {et~al.} 2007, The Astrophysical
  Journal, 657, L105

\bibitem[{Foreman-Mackey(2016)}]{corner}
Foreman-Mackey, D. 2016, The Journal of Open Source Software, 1, 24

\bibitem[{{Foreman-Mackey} {et~al.}(2013){Foreman-Mackey}, {Hogg}, {Lang}, \&
  {Goodman}}]{2013PASP..125..306F}
{Foreman-Mackey}, D., {Hogg}, D.~W., {Lang}, D., \& {Goodman}, J. 2013, \pasp,
  125, 306

\bibitem[{{Fransson} {et~al.}(1996){Fransson}, {Lundqvist}, \&
  {Chevalier}}]{1996ApJ...461..993F}
{Fransson}, C., {Lundqvist}, P., \& {Chevalier}, R.~A. 1996, \apj, 461, 993

\bibitem[{Gall {et~al.}(2015)Gall, Polshaw, Kotak, Jerkstrand, Leibundgut,
  Rabinowitz, Sollerman, Sullivan, Smartt, Anderson, Benetti, Baltay, Feindt,
  Fraser, González-Gaitán, Inserra, Maguire, McKinnon, Valenti, \&
  Young}]{10.1051/0004-6361/201525868}
Gall, E. E.~E., Polshaw, J., Kotak, R., {et~al.} 2015, Astronomy \&
  Astrophysics, 582, A3

\bibitem[{{Grefensetette} {et~al.}(2017){Grefensetette}, {Harrison}, \&
  {Brightman}}]{2017ATel10427....1G}
{Grefensetette}, B., {Harrison}, F., \& {Brightman}, M. 2017, The Astronomer's
  Telegram, 10427, 1

\bibitem[{Heger {et~al.}(2003)Heger, Fryer, Woosley, Langer, \&
  Hartmann}]{10.1086/375341}
Heger, A., Fryer, C.~L., Woosley, S.~E., Langer, N., \& Hartmann, D.~H. 2003,
  The Astrophysical Journal, 591, 288

\bibitem[{Hirschi {et~al.}(2004)Hirschi, Meynet, \&
  Maeder}]{10.1051/0004-6361:20041095}
Hirschi, R., Meynet, G., \& Maeder, A. 2004, Astronomy \& Astrophysics, 425,
  649

\bibitem[{Horesh {et~al.}(2013{\natexlab{a}})Horesh, Kulkarni, Corsi, Frail,
  Cenko, Ben-Ami, Gal-Yam, Yaron, Arcavi, Kasliwal, \&
  Ofek}]{10.1088/0004-637x/778/1/63}
Horesh, A., Kulkarni, S.~R., Corsi, A., {et~al.} 2013{\natexlab{a}}, The
  Astrophysical Journal, 778, 63

\bibitem[{Horesh {et~al.}(2012)Horesh, Kulkarni, Fox, Carpenter, Kasliwal,
  Ofek, Quimby, Gal-Yam, Cenko, Bruyn, Kamble, Wijers, Horst, Kouveliotou,
  Podsiadlowski, Sullivan, Maguire, Howell, Nugent, Gehrels, Law, Poznanski, \&
  Shara}]{10.1088/0004-637x/746/1/21}
Horesh, A., Kulkarni, S.~R., Fox, D.~B., {et~al.} 2012, The Astrophysical
  Journal, 746, 21

\bibitem[{Horesh {et~al.}(2020)Horesh, Sfaradi, Ergon, Barbarino, Sollerman,
  Moldon, Dobie, Schulze, P{\'{e}}rez-Torres, Williams, Fremling, Gal-Yam,
  Kulkarni, O'Brien, Lundqvist, Murphy, Fender, Anand, Belicki, Bellm,
  Coughlin, De, Golkhou, Graham, Green, Hankins, Kasliwal, Kupfer, Laher,
  Masci, Miller, Neill, Ofek, Perrott, Porter, Reiley, Rigault, Rodriguez,
  Rusholme, Shupe, \& Titterington}]{Horesh_2020}
Horesh, A., Sfaradi, I., Ergon, M., {et~al.} 2020, The Astrophysical Journal,
  903, 132

\bibitem[{Horesh {et~al.}(2013{\natexlab{b}})Horesh, Stockdale, Fox, Frail,
  Carpenter, Kulkarni, Ofek, Gal-Yam, Kasliwal, Arcavi, Quimby, Cenko, Nugent,
  Bloom, Law, Poznanski, Gorbikov, Polishook, Yaron, Ryder, Weiler, Bauer, Dyk,
  Immler, Panagia, Pooley, \& Kassim}]{10.1093/mnras/stt1645}
Horesh, A., Stockdale, C., Fox, D.~B., {et~al.} 2013{\natexlab{b}}, Monthly
  Notices of the Royal Astronomical Society, 436, 1258

\bibitem[{{Hosseinzadeh} {et~al.}(2016){Hosseinzadeh}, {Arcavi}, {McCully},
  {Howell}, {Sand}, \& {Valenti}}]{2016ATel.8567....1H}
{Hosseinzadeh}, G., {Arcavi}, I., {McCully}, C., {et~al.} 2016, The
  Astronomer's Telegram, 8567, 1

\bibitem[{Huang {et~al.}(2016)Huang, Wang, Zampieri, Pumo, Arcavi, Brown,
  Graham, Filippenko, Zheng, Hosseinzadeh, Howell, McCully, Rui, Valenti,
  Zhang, Zhang, Zhang, \& Wang}]{10.3847/0004-637x/832/2/139}
Huang, F., Wang, X., Zampieri, L., {et~al.} 2016, The Astrophysical Journal,
  832, 139

\bibitem[{Huang {et~al.}(2015)Huang, Wang, Zhang, Brown, Zampieri, Pumo, Zhang,
  Chen, Mo, \& Zhao}]{10.1088/0004-637x/807/1/59}
Huang, F., Wang, X., Zhang, J., {et~al.} 2015, The Astrophysical Journal, 807,
  59

\bibitem[{Huang {et~al.}(2018)Huang, Wang, Hosseinzadeh, Brown, Mo, Zhang,
  Zhang, Zhang, Howell, Arcavi, McCully, Valenti, Rui, Song, Xiang, Li, Lin, \&
  Wang}]{10.1093/mnras/sty066}
Huang, F., Wang, X.-F., Hosseinzadeh, G., {et~al.} 2018, Monthly Notices of the
  Royal Astronomical Society, 475, 3959

\bibitem[{Hunter(2007)}]{Hunter:2007}
Hunter, J.~D. 2007, Computing In Science \& Engineering, 9, 90

\bibitem[{{Immler} \& {Brown}(2012)}]{2012ATel.3995....1I}
{Immler}, S. \& {Brown}, P.~J. 2012, The Astronomer's Telegram, 3995, 1

\bibitem[{Inserra {et~al.}(2011)Inserra, Turatto, Pastorello, Benetti,
  Cappellaro, Pumo, Zampieri, Agnoletto, Bufano, Botticella, Valle, Rosa,
  Iijima, Spiro, \& Valenti}]{10.1111/j.1365-2966.2011.19128.x}
Inserra, C., Turatto, M., Pastorello, A., {et~al.} 2011, Monthly Notices of the
  Royal Astronomical Society, 417, 261

\bibitem[{Kamble {et~al.}(2016)Kamble, Margutti, Soderberg, Chakraborti,
  Fransson, Chevalier, Powell, Milisavljevic, Parrent, \&
  Bietenholz}]{10.3847/0004-637x/818/2/111}
Kamble, A., Margutti, R., Soderberg, A.~M., {et~al.} 2016, The Astrophysical
  Journal, 818, 111

\bibitem[{Leonard {et~al.}(2002)Leonard, Filippenko, Gates, Li, Eastman, Barth,
  Bus, Chornock, Coil, Frink, Grady, Harris, Malkan, Matheson, Quirrenbach, \&
  Treffers}]{10.1086/324785}
Leonard, D., Filippenko, A., Gates, E., {et~al.} 2002, Publications of the
  Astronomical Society of the Pacific, 114, 35

\bibitem[{{Li} {et~al.}(2005){Li}, {Van Dyk}, {Filippenko}, \&
  {Cuillandre}}]{2005PASP..117..121L}
{Li}, W., {Van Dyk}, S.~D., {Filippenko}, A.~V., \& {Cuillandre}, J.-C. 2005,
  \pasp, 117, 121

\bibitem[{{Lundqvist} \& {Fransson}(1988)}]{1988A&A...192..221L}
{Lundqvist}, P. \& {Fransson}, C. 1988, \aap, 192, 221

\bibitem[{{Margutti} {et~al.}(2014){Margutti}, {Milisavljevic}, {Soderberg},
  {Chornock}, {Zauderer}, {Murase}, {Guidorzi}, {Sanders}, {Kuin}, {Fransson},
  {Levesque}, {Chandra}, {Berger}, {Bianco}, {Brown}, {Challis},
  {Chatzopoulos}, {Cheung}, {Choi}, {Chomiuk}, {Chugai}, {Contreras}, {Drout},
  {Fesen}, {Foley}, {Fong}, {Friedman}, {Gall}, {Gehrels}, {Hjorth}, {Hsiao},
  {Kirshner}, {Im}, {Leloudas}, {Lunnan}, {Marion}, {Martin}, {Morrell},
  {Neugent}, {Omodei}, {Phillips}, {Rest}, {Silverman}, {Strader},
  {Stritzinger}, {Szalai}, {Utterback}, {Vinko}, {Wheeler}, {Arnett},
  {Campana}, {Chevalier}, {Ginsburg}, {Kamble}, {Roming}, {Pritchard}, \&
  {Stringfellow}}]{2014ApJ...780...21M}
{Margutti}, R., {Milisavljevic}, D., {Soderberg}, A.~M., {et~al.} 2014, \apj,
  780, 21

\bibitem[{Margutti {et~al.}(2014)Margutti, Milisavljevic, Soderberg, Guidorzi,
  Morsony, Sanders, Chakraborti, Ray, Kamble, Drout, Parrent, Zauderer, \&
  Chomiuk}]{10.1088/0004-637x/797/2/107}
Margutti, R., Milisavljevic, D., Soderberg, A.~M., {et~al.} 2014, The
  Astrophysical Journal, 797, 107

\bibitem[{{Maund} \& {Smartt}(2005)}]{2005MNRAS.360..288M}
{Maund}, J.~R. \& {Smartt}, S.~J. 2005, \mnras, 360, 288

\bibitem[{{McMullin} {et~al.}(2007){McMullin}, {Waters}, {Schiebel}, {Young},
  \& {Golap}}]{2007ASPC..376..127M}
{McMullin}, J.~P., {Waters}, B., {Schiebel}, D., {Young}, W., \& {Golap}, K.
  2007, in Astronomical Society of the Pacific Conference Series, Vol. 376,
  Astronomical Data Analysis Software and Systems XVI, ed. R.~A. {Shaw},
  F.~{Hill}, \& D.~J. {Bell}, 127

\bibitem[{{Misra} {et~al.}(2007){Misra}, {Pooley}, {Chandra}, {Bhattacharya},
  {Ray}, {Sagar}, \& {Lewin}}]{2007MNRAS.381..280M}
{Misra}, K., {Pooley}, D., {Chandra}, P., {et~al.} 2007, \mnras, 381, 280

\bibitem[{Moriya {et~al.}(2014)Moriya, Maeda, Taddia, Sollerman, Blinnikov, \&
  Sorokina}]{10.1093/mnras/stu163}
Moriya, T.~J., Maeda, K., Taddia, F., {et~al.} 2014, Monthly Notices of the
  Royal Astronomical Society, 439, 2917

\bibitem[{{Nayana} {et~al.}(2018){Nayana}, {Chandra}, \&
  {Ray}}]{2018ApJ...863..163N}
{Nayana}, A.~J., {Chandra}, P., \& {Ray}, A.~K. 2018, \apj, 863, 163

\bibitem[{{Nymark} {et~al.}(2006){Nymark}, {Fransson}, \&
  {Kozma}}]{2006A&A...449..171N}
{Nymark}, T.~K., {Fransson}, C., \& {Kozma}, C. 2006, \aap, 449, 171

\bibitem[{Ofek {et~al.}(2014)Ofek, Sullivan, Shaviv, Steinbok, Arcavi, Gal-Yam,
  Tal, Kulkarni, Nugent, Ben-Ami, Kasliwal, Cenko, Laher, Surace, Bloom,
  Filippenko, Silverman, \& Yaron}]{10.1088/0004-637x/789/2/104}
Ofek, E.~O., Sullivan, M., Shaviv, N.~J., {et~al.} 2014, The Astrophysical
  Journal, 789, 104

\bibitem[{Oliphant(2006)}]{Oliphant:2006wm}
Oliphant, T.~E. 2006, {A Guide to NumPy}, USA

\bibitem[{{Pacholczyk}(1970)}]{1970ranp.book.....P}
{Pacholczyk}, A.~G. 1970, {Radio astrophysics. Nonthermal processes in galactic
  and extragalactic sources} (San Francisco, California, USA: W. H. Freeman)

\bibitem[{Pastorello {et~al.}(2007)Pastorello, Smartt, Mattila, Eldridge,
  Young, Itagaki, Yamaoka, Navasardyan, Valenti, Patat, Agnoletto, Augusteijn,
  Benetti, Cappellaro, Boles, Bonnet-Bidaud, Botticella, Bufano, Cao, Deng,
  Dennefeld, Elias-Rosa, Harutyunyan, Keenan, Iijima, Lorenzi, Mazzali, Meng,
  Nakano, Nielsen, Smoker, Stanishev, Turatto, Xu, \&
  Zampieri}]{10.1038/nature05825}
Pastorello, A., Smartt, S.~J., Mattila, S., {et~al.} 2007, Nature, 447, 829

\bibitem[{Pastorello {et~al.}(2009)Pastorello, Valenti, Zampieri, Navasardyan,
  Taubenberger, Smartt, Arkharov, Bärnbantner, Barwig, Benetti, Birtwhistle,
  Botticella, Cappellaro, Principe, Mille, Rico, Dolci, Elias‐Rosa, Efimova,
  Fiedler, Harutyunyan, Höflich, Kloehr, Larionov, Lorenzi, Maund, Napoleone,
  Ragni, Richmond, Ries, Spiro, Temporin, Turatto, \&
  Wheeler}]{10.1111/j.1365-2966.2009.14505.x}
Pastorello, A., Valenti, S., Zampieri, L., {et~al.} 2009, Monthly Notices of
  the Royal Astronomical Society, 394, 2266

\bibitem[{{Pooley} \& {Lewin}(2002)}]{2002IAUC.8024....2P}
{Pooley}, D. \& {Lewin}, W.~H.~G. 2002, \iaucirc, 8024, 2

\bibitem[{{Pooley} {et~al.}(2002){Pooley}, {Lewin}, {Fox}, {Miller}, {Lacey},
  {Van Dyk}, {Weiler}, {Sramek}, {Filippenko}, {Leonard}, {Immler},
  {Chevalier}, {Fabian}, {Fransson}, \& {Nomoto}}]{2002ApJ...572..932P}
{Pooley}, D., {Lewin}, W. H.~G., {Fox}, D.~W., {et~al.} 2002, \apj, 572, 932

\bibitem[{{Ryder} {et~al.}(2004){Ryder}, {Sadler}, {Subrahmanyan}, {Weiler},
  {Panagia}, \& {Stockdale}}]{2004MNRAS.349.1093R}
{Ryder}, S.~D., {Sadler}, E.~M., {Subrahmanyan}, R., {et~al.} 2004, \mnras,
  349, 1093

\bibitem[{{Schlegel}(2001)}]{2001ApJ...556L..25S}
{Schlegel}, E.~M. 2001, \apjl, 556, L25

\bibitem[{Schlegel(2001)}]{Schlegel_2001}
Schlegel, E.~M. 2001, The Astrophysical Journal, 556, L25

\bibitem[{Smartt(2009)}]{10.1146/annurev-astro-082708-101737}
Smartt, S.~J. 2009, Annual Review of Astronomy and Astrophysics, 47, 63

\bibitem[{Smartt(2015)}]{10.1017/pasa.2015.17}
Smartt, S.~J. 2015, Publications of the Astronomical Society of Australia, 32,
  e016

\bibitem[{Smith(2014)}]{10.1146/annurev-astro-081913-040025}
Smith, N. 2014, Annual Review of Astronomy and Astrophysics, 52, 1

\bibitem[{Smith {et~al.}(2009)Smith, Hinkle, \&
  Ryde}]{10.1088/0004-6256/137/3/3558}
Smith, N., Hinkle, K.~H., \& Ryde, N. 2009, The Astronomical Journal, 137, 3558

\bibitem[{Smith {et~al.}(2010)Smith, Miller, Li, Filippenko, Silverman, Howard,
  Nugent, Marcy, Bloom, Ghez, Lu, Yelda, Bernstein, \&
  Colucci}]{10.1088/0004-6256/139/4/1451}
Smith, N., Miller, A., Li, W., {et~al.} 2010, The Astronomical Journal, 139,
  1451

\bibitem[{Soderberg {et~al.}(2010)Soderberg, Brunthaler, Nakar, Chevalier, \&
  Bietenholz}]{10.1088/0004-637x/725/1/922}
Soderberg, A.~M., Brunthaler, A., Nakar, E., Chevalier, R.~A., \& Bietenholz,
  M.~F. 2010, The Astrophysical Journal, 725, 922

\bibitem[{{Soderberg} {et~al.}(2005){Soderberg}, {Kulkarni}, {Berger},
  {Chevalier}, {Frail}, {Fox}, \& {Walker}}]{2005ApJ...621..908S}
{Soderberg}, A.~M., {Kulkarni}, S.~R., {Berger}, E., {et~al.} 2005, \apj, 621,
  908

\bibitem[{{Soderberg} {et~al.}(2012){Soderberg}, {Margutti}, {Zauderer},
  {Krauss}, {Katz}, {Chomiuk}, {Dittmann}, {Nakar}, {Sakamoto}, {Kawai},
  {Hurley}, {Barthelmy}, {Toizumi}, {Morii}, {Chevalier}, {Gurwell},
  {Petitpas}, {Rupen}, {Alexander}, {Levesque}, {Fransson}, {Brunthaler},
  {Bietenholz}, {Chugai}, {Grindlay}, {Copete}, {Connaughton}, {Briggs},
  {Meegan}, {von Kienlin}, {Zhang}, {Rau}, {Golenetskii}, {Mazets}, \&
  {Cline}}]{2012ApJ...752...78S}
{Soderberg}, A.~M., {Margutti}, R., {Zauderer}, B.~A., {et~al.} 2012, \apj,
  752, 78

\bibitem[{Sorce {et~al.}(2014)Sorce, Tully, Courtois, Jarrett, Neill, \&
  Shaya}]{10.1093/mnras/stu1450}
Sorce, J.~G., Tully, R.~B., Courtois, H.~M., {et~al.} 2014, Monthly Notices of
  the Royal Astronomical Society, 444, 527

\bibitem[{{Stockdale} {et~al.}(2003){Stockdale}, {Weiler}, {Van Dyk}, {Montes},
  {Panagia}, {Sramek}, {Perez-Torres}, \& {Marcaide}}]{2003ApJ...592..900S}
{Stockdale}, C.~J., {Weiler}, K.~W., {Van Dyk}, S.~D., {et~al.} 2003, \apj,
  592, 900

\bibitem[{Strotjohann {et~al.}(2021)Strotjohann, Ofek, Gal-Yam, Bruch, Schulze,
  Shaviv, Sollerman, Filippenko, Yaron, Fremling, Nordin, Kool, Perley, Ho,
  Yang, Yao, Soumagnac, Graham, Barbarino, Tartaglia, De, Goldstein, Cook,
  Brink, Taggart, Yan, Lunnan, Kasliwal, Kulkarni, Nugent, Masci, Rosnet,
  Adams, Andreoni, Bagdasaryan, Bellm, Burdge, Duev, Dugas, Frederick,
  Goldwasser, Hankins, Irani, Karambelkar, Kupfer, Liang, Neill, Porter,
  Riddle, Sharma, Short, Taddia, Tzanidakis, Roestel, Walters, \&
  Zhuang}]{10.3847/1538-4357/abd032}
Strotjohann, N.~L., Ofek, E.~O., Gal-Yam, A., {et~al.} 2021, The Astrophysical
  Journal, 907, 99

\bibitem[{{Szalai} {et~al.}(2019){Szalai}, {Vink{\'o}}, {K{\"o}nyves-T{\'o}th},
  {Nagy}, {Bostroem}, {S{\'a}rneczky}, {Brown}, {Pejcha}, {B{\'o}di}, {Cseh},
  {Cs{\"o}rnyei}, {Dencs}, {Hanyecz}, {Ign{\'a}cz}, {Kalup}, {Kriskovics},
  {Ordasi}, {P{\'a}l}, {Seli}, {S{\'o}dor}, {Szak{\'a}ts}, {Vida}, {Zsidi},
  {Konkoly Team}, {Arcavi}, {Ashall}, {Burke}, {Galbany}, {Hiramatsu},
  {Hosseinzadeh}, {Hsiao}, {Howell}, {McCully}, {Moran}, {Rho}, {Sand},
  {Shahbandeh}, {Valenti}, {Wang}, {Wheeler}, \& {Supernova
  Project}}]{2019ApJ...876...19S}
{Szalai}, T., {Vink{\'o}}, J., {K{\"o}nyves-T{\'o}th}, R., {et~al.} 2019, \apj,
  876, 19

\bibitem[{Utrobin \& Chugai(2019)}]{10.1093/mnras/stz2716}
Utrobin, V.~P. \& Chugai, N.~N. 2019, Monthly Notices of the Royal Astronomical
  Society, 490, 2042

\bibitem[{{Van Dyk} {et~al.}(2003){Van Dyk}, {Li}, \&
  {Filippenko}}]{2003PASP..115....1V}
{Van Dyk}, S.~D., {Li}, W., \& {Filippenko}, A.~V. 2003, \pasp, 115, 1

\bibitem[{{van Loon}(2010)}]{2010ASPC..425..279V}
{van Loon}, J.~T. 2010, in Astronomical Society of the Pacific Conference
  Series, Vol. 425, Hot and Cool: Bridging Gaps in Massive Star Evolution, ed.
  C.~{Leitherer}, P.~D. {Bennett}, P.~W. {Morris}, \& J.~T. {Van Loon}, 279

\bibitem[{Weiler {et~al.}(1991)Weiler, Dyk, Discenna, Panagia, \&
  Sramek}]{10.1086/170571}
Weiler, K.~W., Dyk, S. D.~v., Discenna, J.~L., Panagia, N., \& Sramek, R.~A.
  1991, The Astrophysical Journal, 380, 161

\bibitem[{Weiler {et~al.}(2002)Weiler, Panagia, Montes, \&
  Sramek}]{10.1146/annurev.astro.40.060401.093744}
Weiler, K.~W., Panagia, N., Montes, M.~J., \& Sramek, R.~A. 2002, Annual Review
  of Astronomy and Astrophysics, 40, 387

\bibitem[{{Weiler} {et~al.}(1990){Weiler}, {Panagia}, \&
  {Sramek}}]{1990ApJ...364..611W}
{Weiler}, K.~W., {Panagia}, N., \& {Sramek}, R.~A. 1990, \apj, 364, 611

\bibitem[{Yadav {et~al.}(2014)Yadav, Ray, Chakraborti, Stockdale, Chandra,
  Smith, Roy, Bose, Dwarkadas, Sutaria, \& Pooley}]{10.1088/0004-637x/782/1/30}
Yadav, N., Ray, A., Chakraborti, S., {et~al.} 2014, The Astrophysical Journal,
  782, 30

\bibitem[{{Zheng} \& {Zhang}(2016)}]{2016ATel.8584....1Z}
{Zheng}, X.-M. \& {Zhang}, J.-J. 2016, The Astronomer's Telegram, 8584, 1

\end{thebibliography}

\end{document}